\documentclass[a4paper,11pt]{article}
\pdfoutput=1 % if your are submitting a pdflatex (i.e. if you have
\usepackage{jheppub}
\usepackage[T1]{fontenc} % if needed
\usepackage{longtable}
\usepackage[table]{xcolor}
\usepackage{amsmath}
\usepackage{eqnarray}
\usepackage{mathtools}
\usepackage{amssymb}
\usepackage{subfiles}
\usepackage{slashed}
\usepackage{mwe}
\usepackage[caption=false]{subfig}
\usepackage{eqnarray}

\DeclarePairedDelimiter{\evdel}{\langle}{\rangle}
\newcommand{\ev}{\evdel}

\title{\boldmath Gauged $L_\mu$-$L_\tau$ Model with an Inverse Seesaw Mechanism for Neutrino Masses}

\author{Abhish Dev}

\affiliation[]{Maryland Center for Fundamental Physics (MCFP), University of Maryland,\\College Park, U.S.A}
\emailAdd{adev@umd.edu}

\abstract{In this paper, we propose a $G_{L-R}\times U(1)_{L_\mu-L_\tau}$ gauge-symmetric model where $G_{L-R}$ is the left-right gauge symmetry and $L_i$ is the $i-$flavor lepton number. We use the spontaneous breaking (SSB) of $U(1)_{L_\mu-L_\tau}$ to resolve two discrepancies in the standard model: discrepancy in the measured muon anomalous magnetic moment and the small masses of left-handed neutrinos and its oscillations. The massive neutral gauge boson, $Z_{\mu\tau}$, arising from the SSB can provide additional contributions to the muon anomalous magnetic moment. In order to explain neutrino masses, we employ the low-energy inverse seesaw mechanism by adding three $G_{L-R}$ singlet fermions, $S_{e,\mu,\tau}$. The light neutrino mass matrix from the inverse seesaw formula has a specific two-zero texture pattern referred in the literature as the Type-C two-zero texture, due to the $U(1)_{L_\mu-L_\tau}$ symmetry. This allows us to predict the values of the CP-violating Dirac phase, Majorana phases, and the absolute value of light neutrino masses in terms of the precisely measured mixing angles and mass squared differences. The model accommodates a quasi-degenerate spectrum of neutrino masses with inverted ordering. The calculated best-fit value of $\delta_{CP}$ surprisingly matches with the current experimentally measured best-fit value of $\delta_{CP}$ in \cite{Capozzi:2017ipn}. At $1\sigma$, the measured value of $\delta_{CP}$ favors a $\theta_{23}>\pi/4$. At $1\sigma$, most of the parameter space is within the cosmological bound on the sum of neutrino mass and the bound on the effective Majorana mass from neutrinoless beta decay.}
\begin{document} 
\maketitle

\section{Introduction}	
The standard model (SM) can explain most of the observed particle phenomena as we know today. However, there is evidence of new physics in unexplained phenomena such as neutrino masses and oscillations, dark matter, and discrepancies in the experimentally measured value of muon anomalous magnetic moment. One of the ways to build on SM is to extend its gauge structure with the simplest extension being that of an Abelian $U(1)_{X}$ symmetry. Standard model, in addition to its gauge symmetry, has some accidental global $U(1)$ symmetries like the baryon number $(B)$, flavor lepton numbers ($L_e,~L_\mu,$ and $L_\tau$), and the total lepton number ($L=L_e+L_\mu+L_\tau$). It is possible to promote some combinations of these symmetries to local gauge symmetries in an anomaly-free manner. These combinations are $L_e-L_\mu,~L_\mu-L_\tau$, and $L_e-L_\tau$ where the anomalies cancel among the different generations of leptons\cite{Foot:1990mn,He:1990pn,Foot:1994vd,Ma:2001md}.
Out of these three combinations, the most popular one for model building is $L_\mu-L_\tau$  which has been used in  \cite{Xing:2015fdg, Brignole:2004ah, Mohapatra:2005yu, Xing:2006xa,
Adhikary:2006rf, Kitabayashi:2005fc, Baek:2008nz, He:2011kn, Ge:2010js,
EstebanPretel:2007yq, Heeck:2011wj, Joshipura:2009tg, Fuki:2006ag, Aizawa:2005yy,
Adhikary:2009kz, Grimus:2012hu, Grimus:2005jk, Rodejohann:2005ru,
Haba:2006hc, Joshipura:2009fu, Xing:2010ez,
Bandyopadhyay:2009xa, Grimus:2006jz, Fukuyama:1997ky,
Araki:2010kq, Feng:2012jn, Kile:2014jea, Kim:2015fpa,
Zhao:2016orh, Patra:2016shz}. The  $U(1)_{L_\mu-L_\tau}$ gauge symmetry has been used to explain leptogenesis\cite{Chun:2007vh,Ota:2006xr,Adhikary:2006rf}, dark matter\cite{Baek:2017sew,Biswas:2016yjr,Altmannshofer:2016jzy,Biswas:2016yan,Baek:2015fea,Das:2013jca}, baryon asymmetry, cosmic neutrino spectrum, muon (g-2)\cite{Biswas:2016yjr,Biswas:2016yan,Patra:2016shz,Altmannshofer:2016oaq,Baek:2015fea,Araki:2014ona,Harigaya:2013twa,Ma:2001md}, neutrino masses\cite{Chen:2017gvf,Asai:2017ryy,Biswas:2016yjr,Biswas:2016yan, Chun:2007vh,Adhikary:2006rf,Rodejohann:2005ru,Ma:2001md}, and B anomalies at the LHCb and Belle experiments\cite{Baek:2017sew,Chen:2017usq,Altmannshofer:2016jzy,Crivellin:2015mga}. 

In this paper, we propose a $U(1)_{L_\mu-L_\tau}$ gauge model which can explain both the muon anomalous magnetic moment and the small neutrino masses. This strategy is suitable because a $U(1)_{L_\mu-L_\tau}$ gauge model automatically satisfies $\mu-\tau$ permutation symmetry in the leading order thereby naturally accommodating a maximal $\theta_{23}$ and a vanishing $\theta_{13}$. The $U(1)_{L_\mu-L_\tau}$ gauge symmetry is not an exact symmetry and must be broken due to the non-vanishing VEV of an ${L_\mu-L_\tau}$ charged Higgs field. This spontaneous symmetry breaking (SSB) of the $U(1)_{L_\mu-L_\tau}$ then provides the necessary mass splitting and deviations from maximal $\theta_{23}$ and vanishing $\theta_{13}$. Lepton flavor violations in $L_{\mu}-L_{\tau}$ models is analyzed in \cite{Heeck:2014qea}. The neutral gauge boson arising from the SSB, $Z_{\mu\tau}$, can provide additional contributions to rectify the discrepancy between the theoretically calculated and the experimentally measured values of muon $g-2$. $Z_{\mu\tau}$ will not couple to the electron or quarks making it insular to the constraints from hadron and lepton colliders. The scope of detecting $Z_{\mu\tau}$ at the BELLE II
\cite{Chen:2017cic,
Araki:2017wyg,Kaneta:2016uyt}, NA64 experiment at CERN\cite{Gninenko:2014pea,
Banerjee:2017hhz} , and neutrino beam experiments \cite{Kaneta:2016uyt} has been widely investigated in previous studies.

A series of CERN experiments and E821 experiment at the Brookhaven National Laboratory(BNL) have reported an anomaly between the experimentally measured value of muon (g-2) and its SM prediction. The strongest constraint in the $g_{\mu\tau}-m_{Z_{\mu\tau}}$ parameter space comes from the neutrino trident cross section\cite{Altmannshofer:2014pba} which is measured in the CHARM II\cite{Geiregat:1990gz} and CCFR experiments\cite{Mishra:1991bv}. Constraints on $Z_{\mu\tau}$ from LHC are analyzed in \cite{Elahi:2017ppe, Elahi:2015vzh}. In section \ref{MAMM}, we show that accommodating muon g-2 strongly limits the breaking scale of $U(1)_{L_\mu-L_\tau}$ to be in the range 12-800 GeV.

One of the long-standing issues in physics is the existence of light left-handed neutrinos which the SM predicts to be massless. Neutrino oscillation experiments and cosmological observations have confirmed the existence of three flavors of neutrinos which mix with the values of oscillation parameters given in the table.\ref{data} which are the most recently updated values taken from \cite{Capozzi:2017ipn}. The small neutrino masses can be elegantly realized by using seesaw mechanisms. In the type-I, -II, and -III seesaw models, the total lepton number is spontaneously broken at a high scale, $\Lambda$, which generates sub-eV masses via the seesaw formula $m_\nu=\frac{v^2}{\Lambda}$, where $v$ is the electroweak scale. However, this also pushes $\Lambda$ beyond the multi-TeV scales making the new physics inaccessible at the LHC. Using the spontaneous symmetry breaking of $U(1)_{L_\mu-L_\tau}$ to explain both muon anomalous magnetic moment and neutrino masses has been realized in various models in the context of Type I seesaw mechanism\cite{Biswas:2016yjr,Biswas:2016yan,Patra:2016shz}. Since $U(1)_{L_\mu-L_\tau}$ has to be broken in the GeV scale, it is natural to prefer a low energy seesaw mechanism like the inverse seesaw mechanism over type-I, -II, and -III seesaw mechanisms.

The inverse seesaw mechanism often referred in the literature as the double seesaw mechanism is a low-energy seesaw mechanism in which low energy gauge singlet fermions, $S$, are added in addition to the right-handed neutrinos, $N_R$, to obtain sub-eV left-handed neutrinos, $\nu_L$\cite{Mohapatra:1986aw, Mohapatra:1986bd,GonzalezGarcia:1988rw,
Deppisch:2004fa}. In general, the number of sterile fermions can be different from the number of neutrino species. In the basis $(\nu_L,~N^c_R,~S)$, the lepton mass matrix takes the following peculiar form.
\begin{equation}\label{M}
M_\nu = \left(
\begin{array}{ccc}
 0 & m_D & 0 \\
 m_D & 0 & m_N \\
 0 & m_N & \mu  \\
\end{array}
\right).
\end{equation}
where the matrices $M_N$, $m_D$, and $\mu$ are in general complex. In the hierarchy $\mu\ll m_D\ll M$, the effective light neutrino mass matrix can be approximated as
\begin{equation}\label{inverse}
m_\nu=m_D^T (m_N^T)^{-1}\mu m_N m_D.
\end{equation}
This is the inverse seesaw formula. As opposed to the Type-I seesaw mechanism, it is possible to have the right-handed neutrinos in the  TeV scale thereby providing new physics which is testable at the LHC.

In SM, it is not possible to achieve the peculiar pattern in eq.(\ref{M}). The zero in $(\nu,S)$ positions of $M_\nu$ is usually a result of additional symmetries beyond the SM gauge symmetry. An elegant realization of this peculiar form in eq.(\ref{M}) can be found in the well-studied left-right symmetric (L-R) models\cite{Pati:1974yy,Mohapatra:1974hk,Senjanovic:1975rk,Senjanovic:1978ev,
Mohapatra:1980yp,Lim:1981kv} as pointed out in \cite{Dev:2009aw}. In this class of models, the SM gauge symmetry is enhanced to $SU(3)_C \times SU(2)_L\times SU(2)_R\times U(1)_{B-L}$ where $B$ and $L$ are baryon and lepton numbers respectively. An attractive feature of these models is the replacement of the ad-hoc hyper charge, $Y$, of the SM with the well-motivated $B-L$ quantum number. Analogous to the left-handed fermion $SU(2)_L$ doublets in the SM, right-handed fermions are $SU(2)_R$ doublets in the L-R models. The L-R lagrangian preserves parity symmetry which is spontaneously broken at a scale well above the electroweak scale. L-R models also have a richer Higgs sector to enable these symmetry breaking patterns. The quark and charged lepton masses are obtained by introducing an $SU(2)_L\times SU(2)_R$ bi-doublet $\phi$. In the minimal L-R model with gauge triplet Higgs, light neutrino masses are obtained by a combination of Type-I and Type-II seesaw mechanisms.

Putting all these ingredients together, we get an $SU(3)_C\times SU(2)_L\times SU(2)_R \times U(1)_{B-L}\times U(1)_{L_\mu-L_\tau}$ non-Abelian gauge theory which spontaneously breaks to $SU(3)_C\times U(1)_{em}$. We employ an L-R singlet but $U(1)_{L_\mu-L_\tau}$ charged Higgs, $\delta$, to break the $U(1)_{L_\mu-L_\tau}$  completely. An $SU(2)_R$ doublet Higgs, $\chi_R$, spontaneously breaks $G_{L-R}$ to $G_{SM}$ and an $SU(2)_L\times SU(2)_R$ bi-doublet Higgs, $\phi$, which carries out the further breaking of $G_{SM}$ to $SU(3)_C\times U(1)_{em}$ at the electroweak scale. In the fermionic sector, we add three right-handed neutrinos $(N_e, N_\mu, N_{\tau})$ and three L-R singlet fermions ($S_e,S_\mu,S_\tau$) where the subscript denotes their respective flavor lepton number. After SSB, the mass matrix in the basis $(\nu_L,N_R,S)$ takes the desired form in eq.(\ref{M}) and the inverse seesaw formula can be used to calculate the approximate light neutrino masses. A parameter scan reveals that this model can only accommodate the inverted ordering of light neutrino masses with a quasi-degenerate hierarchy. In sections \ref{3} and \ref{5}, we will discuss the model and the lepton masses arising from this model in detail.

The $U(1)_{L_\mu-L_\tau}$ symmetry restricts the form of Majorana and Dirac mass matrices $m_D$, $m_N$, and $\mu$. $m_D$ and $m_N$ are restricted to be diagonal matrices whereas the matrix $\mu$ has zeroes in its $(\mu,\mu)$ and $(\tau,\tau)$ entries. The approximate light neutrino mass matrix $m_\nu$ calculated using the inverse seesaw formula, therefore, also has zeroes in its $(\mu,\mu)$ and $(\tau,\tau)$ positions. This is one of the seven two-zero neutrino mass textures which can accommodate the current neutrino oscillation data, referred in the literature as the type C pattern. These patterns are studied in \cite{Berger:2000zj,Xing:2002ta,Kageyama:2002zw,Xing:2002ap}. In the context of $U(1)_{L_\mu-L_\tau}$ models, the type-C two-zero pattern is also realized in models where neutrino masses are radiatively generated \cite{Baek:2015mna,
Lee:2017ekw}. The two zeroes in $m_\nu$ lead to four real relations that relate the precisely measured values of the mixing angles ($\theta_{13},~\theta_{12},~\text{and}~\theta_{23}$) and the mass-squared differences ($\Delta m^2_{21}$ and $\Delta m^2_{}$) to  predict the CP-violating Dirac phase ($\delta_{CP}$), the two Majorana phases ($\alpha_{1,2}$), and the absolute values of the neutrino masses, $m_i$. The effective Majorana mass of neutrinoless double beta decay (<$m_{\beta\beta}$>) and the sum of neutrino masses ($\sum m_i$) can be then calculated and compared with the constraints from neutrinoless beta decay and cosmology respectively. In section \ref{6}, it is shown that the major chunk of the parameter space is well within the current experimental bounds. As expected from the parameter scan, this analysis agrees with the quasi-degenerate, inverted ordering for neutrino mass spectrum and a prediction of $\delta_{CP}$ close to the present central value. In contrast, the models based on Type-I seesaw mechanism\cite{Asai:2017ryy,Biswas:2016yjr,Biswas:2016yan} satisfy the type-C pattern of two-zero minor textures with vanishing $(\mu,\mu)$ and $(\tau,\tau)$ entries in the inverse of neutrino mass matrix. The predictions of this texture is analyzed in detail in \cite{Asai:2017ryy}.

\section{The $G_{L-R}\times U(1)_{L_\mu-L_\tau}$ Model}\label{3}

In this section, we will discuss the $G_{L-R}\times U(1)_{L_\mu-L_\tau}$ gauge-symmetric model that we motivated in the introduction. 

In addition to the quarks and leptons in the L-R models, the fermionic sector contains three L-R singlet fermions: $S_e,~S_\mu,~\text{and}~S_\tau$ where the subscript indicates their flavor lepton number. The Higgs sector has the usual bi-doublet Higgs, $\phi$, which is the L-R analog of the SM Higgs doublet. The minimal L-R model has  $SU(2)_{L,R}$ triplet Higgs fields $\Delta_{L,R}$ with a $B-L$ charge of 2 which generate small neutrino masses through a combination of Type-I and Type-II seesaw mechanisms. In our model, we include $SU(2)_{R}$ doublet $\chi_{R}$ with a B-L charge of $-1$ instead of the triplets to get the matrix $m_N$ in eq.(\ref{M}). The  triplets must be avoided to prevent Majorana masses for $\nu_L$ and $N_R$ in eq.(\ref{M}). The gauge group $U(1)_{L_\mu-L_\tau}$ is completely broken by an L-R singlet Higgs $\delta$. Since $\delta$ is an L-R singlet, the $U(1)_{L_\mu-L_\tau}$ breaking scale is independent of both the electroweak and parity breaking scales. We choose $\delta$ to have an $L_\mu-L_\tau$ charge of $\pm 1$ with both being identical choices. A choice of $\pm 2$ would lead to a block-diagonal neutrino mass matrix which cannot generate the desired oscillation patterns. The fermions and scalars in our models and their assignments under various irreps of the gauge group are given in Table.\ref{assign}.
\begin{table}
\centering
\begin{tabular}{c c c c c c}

\hline\hline
Fields& Content& $SU(2)_L$ & $SU(2)_R$&B-L&$L_\mu-L_\tau$\\
\hline
\\
 \hspace {0.35cm} $L_{eL}$ & $\left( \begin{array}{c} \nu_e \\
    l_e\end{array} \right)_L$ & \quad \underline{2} & \quad \underline{1} & \quad -1 & \quad 0 \\
    
    \hspace {0.5cm} $L_{eR}$ & $\left( \begin{array}{c} \nu_e \\
    l_e\end{array} \right)_R$ & \quad \underline{1} & \quad \underline{2} & \quad -1 & \quad 0 \\
    
    \hspace {0.35cm} $L_{\mu L}$ & $\left( \begin{array}{c} \nu_\mu \\    
    l_\mu\end{array} \right)_L$ & \quad \underline{2} & \quad \underline{1} & \quad -1 & \quad 1 \\
    
    \hspace {0.5cm} $L_{\mu R}$ & $\left( \begin{array}{c} \nu_\mu \\
    l_\mu\end{array} \right)_R$ & \quad \underline{1} & \quad \underline{2} & \quad -1 & \quad 1 \\
    
    \hspace {0.35cm} $L_{\tau L}$ & $\left( \begin{array}{c} \nu_\tau \\    
    l_\tau\end{array} \right)_L$ & \quad \underline{1} & \quad \underline{2} & \quad -1 & \quad -1 \\
    
    \hspace {0.5cm} $L_{\tau R}$ & $\left( \begin{array}{c} \nu_\tau \\
    l_\tau\end{array} \right)_R$ & \quad \underline{2} & \quad \underline{1} & \quad -1 & \quad -1 \\
    \\    
    \hline
    \\
    \hspace {0.5cm} $S_{e}$ & $S_{e}$ & \quad \underline{1} & \quad \underline{1} & \quad 0 & \quad 0 \\
    
    \hspace {0.5cm} $S_{\mu}$ & $S_{\mu}$ & \quad \underline{1} & \quad \underline{1} & \quad 0 & \quad 1 \\
    
    \hspace {0.5cm} $S_{\tau}$ & $S_{\tau}$ & \quad \underline{1} & \quad \underline{1} & \quad 0 & \quad -1 \\
    \\
\hline
\\
\hspace {0.7cm} $\phi$ & $\displaystyle{
\left( \begin{array}{cc}
\phi^{0}_{1} & \phi^{+}_{1}\\
\phi^{-}_{2} & \phi^{0}_{2} \end{array} \right)}$ & \quad \underline{2} & \quad \underline{2} & \quad 0 & \quad 0 \\

\hspace {0.5cm} $\chi_R$ & $\left(\begin{array}{c} \chi ^0 \\ \chi ^- \\\end{array}\right)$  & \quad \underline{1} & \quad \underline{2}& \quad -1 & \quad 0 \\

 \hspace {0.5cm} $\delta$ & $\delta$ & \quad \underline{1} & \quad \underline{1} & \quad 0 & \quad 1 \\
\hline\hline

 %\hline%-----------------------------------
\end{tabular}
\caption{The field content and the irrep-assignments of lepton and Higgs fields under $SU(2)_L\times SU(2)_R \times U(1)_{B-L}\times U(1)_{L_\mu-L_\tau}$.}
\label{assign}
\end{table}

The full gauge invariant Lagrangian can be decomposed as
\begin{equation}
\mathcal{L}=\mathcal{L}^{gauge}+\mathcal{L}^{fermion}+\mathcal{L}^{higgs} +\mathcal{L}^{yukawa}-\mathcal{V}(\phi,\chi,\delta)
\end{equation}
where $\mathcal{L}^{gauge}$, $\mathcal{L}^{fermion}$, and $\mathcal{L}^{higgs}$ are the canonically gauge-invariant kinetic terms for the gauge, fermionic, and Higgs sectors respectively. $\mathcal{L}^{yukawa}$ contains the Yukawa interaction terms between the Higgs fields and fermions. $\mathcal{V}(\phi,\chi,\delta)$ is the most general gauge invariant Higgs ponential involving the fields $\phi,~\chi_R,~\text{and}~\delta$ given in Appendix \ref{HP}.

\begin{equation}
\mathcal{L}^{gauge}=\mathcal{L}^{gauge}_{L-R} + B'^{\alpha\beta}B'_{\alpha\beta}
\end{equation}
where $\mathcal{L}^{gauge}_{L-R}$ is gauge-kinetic part of the L-R lagrangian and $B'$ is the gauge boson field associated with $U(1)_{L_\mu-L_\tau}$. For the sake of simplicity, we ignore the possible kinetic mixing term between the Abelian gauge boson fields $B$ and $B'$.
\begin{equation}
\mathcal{L}^{fermion}=\mathcal{L}^{fermion}_{L-R} + \sum_{e,\mu,\tau}~\bar{L}_i(g_{\mu\tau}Q_{\mu\tau}(L_i)\slashed{Z}_{\mu\tau})L_i
+\sum_{e,\mu,\tau}\bar{S}_i(i\slashed{\partial}+g_{\mu\tau}Q_{\mu\tau}(S_i)\slashed{Z}_{\mu\tau} )S_i
\end{equation}
where $Q_{\mu\tau}(X)$ is the $L_\mu-L_\tau$ charge of the field X.

The gauge invariant $\mathcal{L}^{yukawa}$ containing all fermion-scalar interactions is given as

\begin{equation}\label{yukawa}
\begin{aligned}
\mathcal{L}^{yukawa}&=&(h_{Qij} \bar{Q}_{iL}\phi Q_{jR} + \tilde{h}_{Qij} \bar{Q}_{jL}\tilde{\phi} Q_{iR})+(h_{Li} \bar{L}_{iL}\phi L_{iR}
+ \tilde{h}_{Li} \bar{L}_{iL}\tilde{\phi} L_{iR})\\ &&+ f_i \bar{L}_{iR} \chi_R S_i
+ \mu_{ee} S_e S_e  + \mu_{\mu \tau} S_\mu S_\tau + y_{e\mu}S_e S_\mu \delta^*+ + y_{e\tau} S_e S_\tau \delta,
\end{aligned}
\end{equation}
where $i$ and $j$ are the generation indices and $\tilde{\phi}=-i \tau_2 \phi^* \tau_2 i$.

As in SM, we can use the non-vanishing VEV of the neutral Higgs fields to spontaneously break the gauge symmetry. The VEV of $\chi^0$, $v$, breaks the parity symmetry at a multi-TeV scale and the symmetry $SU(2)_R\times U(1)_{B-L}$ is broken into $U(1)_Y$. Similarly, $\phi^0_1$ and $\phi^2_0$ also get non-zero VEV.  For simplicity, we take the VEV of $\phi^2_0$ to be zero. The VEV of $\phi^0_1$ then enables the electroweak symmetry breaking. The Abelian gauge group $U(1)_{L_\mu-L_\tau}$ is spontaneously broken completely by the VEV of $\delta$. The VEVs of the Higgs fields in this model can be summarized as
\begin{equation}
\ev{\phi}=\left(
\begin{array}{cc}
 \kappa  & 0 \\
 0 & 0 \\
\end{array}
\right),~ \ev{\chi}=\left(
\begin{array}{c}
 v \\
 0 \\
\end{array}
\right),~ \ev{\delta}=\delta.
\end{equation}
The VEVs in general can be complex but all phases can be absorbed away by gauge rotations associated with the generators $T_{3R}$, $T_{3L}$, $B-L$, and $L_\mu-L_\tau$ which commute with the  electric charge $Q$. The gauge symmetry is broken to the remnant $U(1)_Q$ after SSB and the Gellman-Nishijima formula for the SM gets modified as
\begin{equation}
Q=T_{3L}+T_{3R} + \frac{(B-L)}{2}.
\end{equation}
The canonical Higgs kinetic terms in $\mathcal{L}^{higgs}$ give rise to mass terms for gauge bosons after SSB. The gauge boson masses and mass eigen states are given in Appendix \ref{ES}.

%%%%%%%%%%%%%%%%%%%%%%%%%%%%%%%%%%%%%%%%%%%%%%%%%%%%%%%%%%%%%%%%%%%%%%%%%%%%%%%%%%%%%%%%%%%%%%%%%%%%%%%
\section{Muon anomalous magnetic moment}\label{MAMM}
\begin{figure}
\includegraphics[scale=0.3]{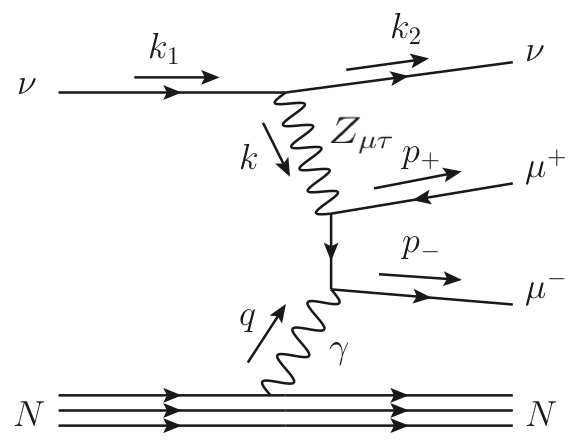}
~~~~~~~~~~~~~~~~~~~~~~~~~~
\includegraphics[scale=0.23]{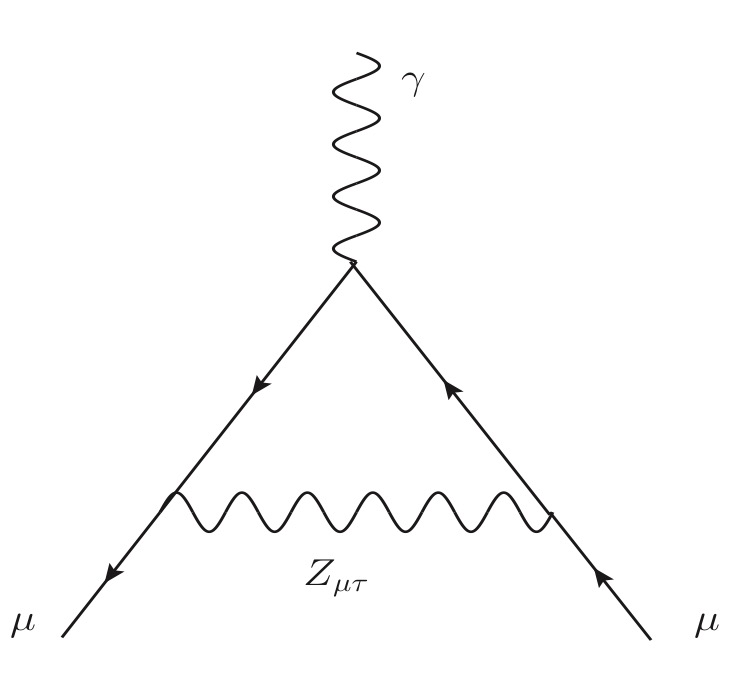}
\caption{(i) The leading order diagram to neutrino trident process mediated by $Z_{\mu\tau}$. (left) There is an additional diagram with $\mu^+$ and $\mu^-$ lines exchanged. (ii) The leading order contribution to muon (g-2) from the $Z_{\mu\tau}$ boson in this model. (right)}
\label{zut}
\end{figure}

The intrinsic spin magnetic moment of a fermion is defined as
\begin{equation}
\vec{\mu}=\frac{g Q e}{2 m}\frac{\vec{\sigma}}{2},
\end{equation}
where $g$ is the gyromagnetic ratio which is predicted to be $2$ for free fermions by the Dirac equation.
Loop corrections to this value can arise from the interactions in a QFT which can give rise to the anomalous magnetic moment defined as
 \begin{equation}
 a_\mu=\frac{g-2}{2}.
 \end{equation}
 Calculation of lepton anomalous magnetic moments and its close agreement with experiments were one of the first major successes of QED. In the standard model, contribution to muon g-2 comes from various sectors such as QED, electroweak, and hadronic vacuum polarization effects. The non-perturbative strong interaction effects to muon g-2 is indirectly determined using the hadronic crossection in electron-positron annihilation thereby introducing uncertainties in the calculated value. This calculation is done in \cite{Jegerlehner:2009ry} which reports a value of $a_\mu^{th}=1.1659179090(65)\times10^{-3}$. The anomalous magnetic moment of muon has been experimentally measured with high precision first in a series of CERN experiments and then by the E821 experiment at Brookhaven National Laboratory (BNL). The present average experimental value stands at  $a_\mu^{exp}=1.16592080(63)\times 10^{-3}$\cite{Bennett:2004pv}. 
 This discrepancy,
 \begin{equation}\label{MAM}
 \Delta a_\mu=a^{exp}_\mu-a^{th}_\mu=(29.0\pm 9.0)\times 10^{-10},
 \end{equation}
 
 is one of the strongest established discrepancy with the standard model at $3.2\sigma$ significance. This has been an inspiration for a lot of model building beyond the SM. A lepton flavor violating $Z'$ was used to account for this discrepancy in \cite{Altmannshofer:2016brv}. The $U(1)_{L_\mu-L_\tau}$ gauge boson couples to the muon and can give rise to additional contributions in Fig.\ref{zut}(ii). This contribution is calculated in \cite{Gninenko:2001hx,Baek:2001kca} to be
 \begin{equation}
 \Delta a_\mu=\frac{g^2_{\mu\tau}}{8\pi^2}\int_{0}^{1}dx\frac{2x(1-x)^2}{(1-x)^2+rx},
 \end{equation} 
 where the dimensionless parameter $r$ is defined as the square of the ratio of $Z_{\mu\tau}$ mass to the mass of muon ($(M_{Z_{\mu\tau}}/m_\mu)^2$). The discrepancy in the value of muon g-2 is accommodated by a wide range of values of $g_{\mu\tau}$ and $m_{Z_{\mu\tau}}$. When $m_{Z_{\mu\tau}}$ falls below $2m_\mu$, $Z_{\mu\tau}$ primarily decays into neutrinos and this region is mostly eliminated by neutrino interactions\cite{Harnik:2012ni,Bilmis:2015lja}. The values of $g_{\mu\tau}>10^{-3}$ is severely constrained primarily by the neutrino trident cross section in  Fig.\ref{zut}(i)\cite{Altmannshofer:2014pba} which has been measured in CHARM-II\cite{Geiregat:1990gz} and CCFR \cite{Mishra:1991bv} experiments.
 
Together, these constraints limit the region in $g_{\mu\tau}-m_{Z_{\mu\tau}}$ plane that can accommodate the necessary $\Delta a_\mu$. The value of $g_{\mu\tau}\approx 5\times 10^{-4}$ and  $3$ MeV<$m_{Z_{\mu\tau}}<200$ MeV can resolve the current discrepancy in the muon anomalous magnetic moment ($\Delta a_\mu$). From eq.(\ref{gaugebosonmasses}), this limits the value of VEV $\delta$ to be between 12-800 GeV. In the following section, we will use this range of values for $\delta$ to fit neutrino masses and oscillation parameters.

 %%%%%%%%%%%%%%%%%%%%%%%%%%%%%%%%%%%%%%%%%%%%%%%%%%%%%%%%%%%%%%

\section{Neutrino masses and oscillations}\label{5}

Fermion masses are generated after SSB from the Higgs-fermion interaction terms, $\mathcal{L}^{yukawa}$, in eq.(\ref{yukawa}). The quark masses are given as
\begin{eqnarray}
M_U=U_{u} \text{diag}\{m_u,m_c,m_t\} U^\dagger_{u}= h_Q \kappa,\\\notag
M_D=U_{d} \text{diag}\{m_d,m_s,m_b\} U^\dagger_{d}= \tilde{h}_Q \kappa,
\end{eqnarray}
where $h_Q$ and $\tilde{h}_Q$ are in general $3\times 3$ hermitian matrices which are diagonalized by unitary transformations with $U_{u}$ and $U_{d}$ matrices respectively. The mixing matrix in the quark sector is then given by $U_{CKM}=U_d^\dagger U_u$.

In the leptonic sector, the charged lepton mass matrix is diagonal and the masses are given by
\begin{equation}
M_l=h_L \kappa=\text{diag}\{m_e,m_\mu,m_\tau\},
\end{equation}
where $h_L$ is a real diagonal $3\times 3$ matrix as the phases can be absorbed by redefining the charged lepton fields. The Yukawa terms for the rest of the fermions are coupled. In the $(\nu_L,N_R,S)$ basis, the $9\times 9$ mass matrix, $M_\nu$, has the form in eq.(\ref{M}) with the $3\times3$ mass matrices $m_D,~m_N,$ and $\mu$ given by
\begin{eqnarray}\notag
m_D&=&\tilde{h}_L \kappa,~~ 
m_N=f v,\\
\mu~&=&\left(\begin{array}{ccc}\label{block}
\mu_{ee}&y_{e\mu}\delta &y_{e\tau}\delta\\
y_{e\mu}\delta&0&\mu_{\mu\tau}\\
y_{e\tau}\delta&\mu_{\mu\tau} &0\end{array}\right).
\end{eqnarray}
The parameters appearing in eq.(\ref{block}) can  be complex but all the phases can be rotated away by redefining the fermion fields except for one which has to appear in one of the $\mu_{ij}$s. We will keep the phase in $\mu_{\mu \tau}$ by redefining it as $\mu_{\mu \tau}=\mu_{\mu\tau}e^{i \phi}$. Using the inverse seesaw formula in eq.(\ref{inverse}), the $3\times 3$ light neutrino mass matrix can now be approximated as
\begin{eqnarray}\label{eq1}
m_\nu &\approx m_D^T (m_N^T)^{-1}\mu m_N m_D
      &=\left(
\begin{array}{ccc}
 \mu_{ee} \alpha _1^2 & y_{e\mu} \delta  \alpha _1 \alpha _2 & y_{e\tau} \delta  \alpha _1 \alpha _3 \\
y_{e\mu} \delta  \alpha _1 \alpha _2 & 0 & \alpha _2 \alpha _3 \mu_{\mu\tau} e^{i \phi} \\
y_{e\tau} \delta  \alpha _1 \alpha _3 & \alpha _2 \alpha _3 \mu_{\mu\tau} e^{i \phi}& 0 \\
\end{array}
\right),
\end{eqnarray}
where $\alpha_i=\tilde{h}_{Li}\kappa/f_iv$.

The light neutrino mass matrix is complex symmetric and is diagonalized by the $U_{PMNS}$ matrix such that $\text{diag}\{m_1,m_2,m_3\}=U_{PMNS}^\dagger m_\nu U_{PMNS}^*$. The matrix in eq.(\ref{eq1}) is constrained by the experimental values of neutrino mass-squared differences, $\Delta m^2_{21}$ and $\Delta m^2_{31}$, and the three mixing angles $\theta_{12},~\theta_{13},$ and $\theta_{23}$ as tabulated in Table.\ref{data}. These are the most updated values as given in \cite{Capozzi:2017ipn}. Additional constraints on CP-violating Dirac phase, $\delta_{CP}$, sum of masses, $\Sigma m_i$, and effective mass of neutrinoless double beta decay, $<m_{\beta\beta}>$, are discussed in detail in section \ref{6}.
\begin{figure}
\hskip-2.0cm\includegraphics[scale=0.35]{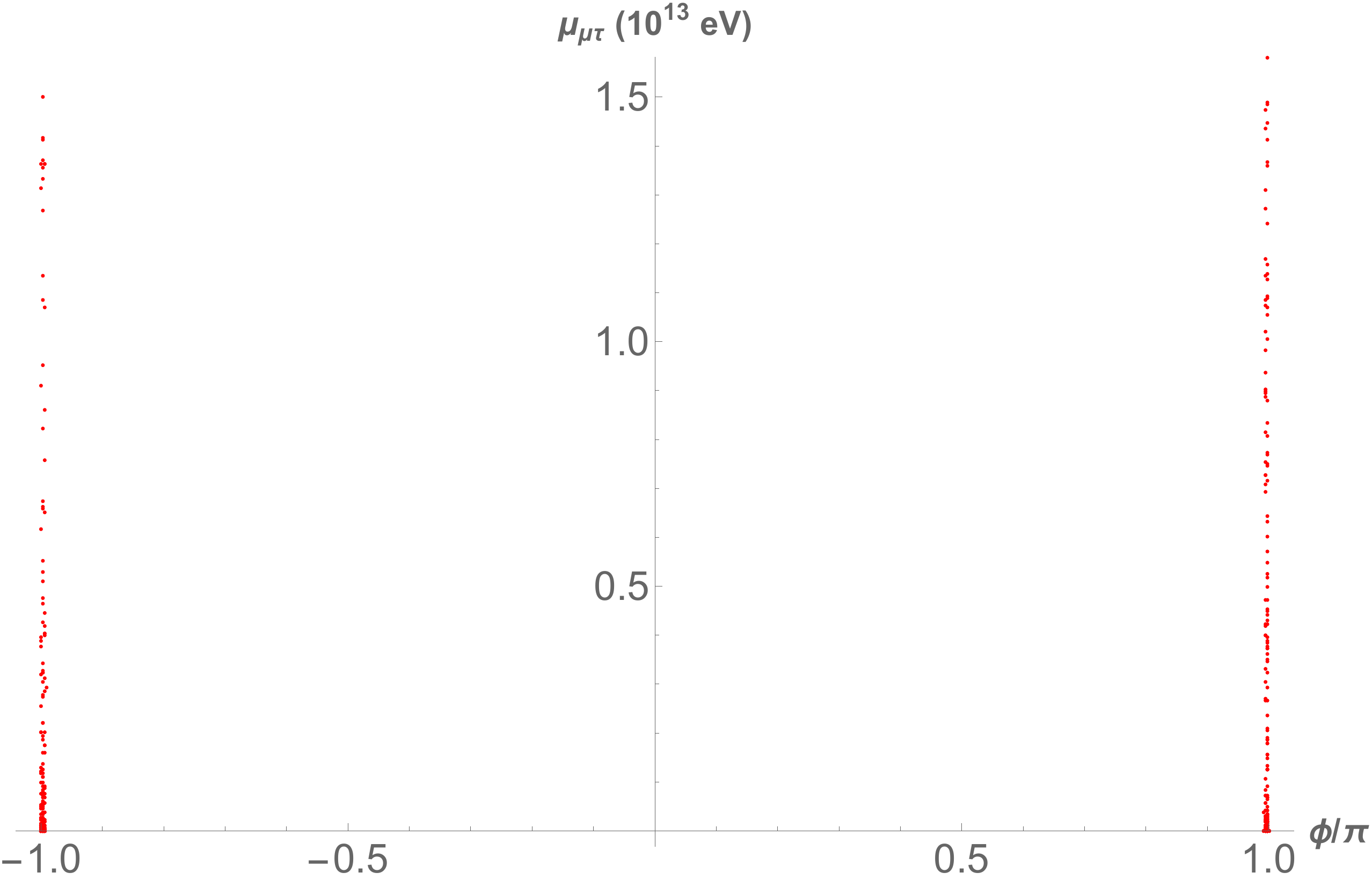}
\includegraphics[scale=0.44]{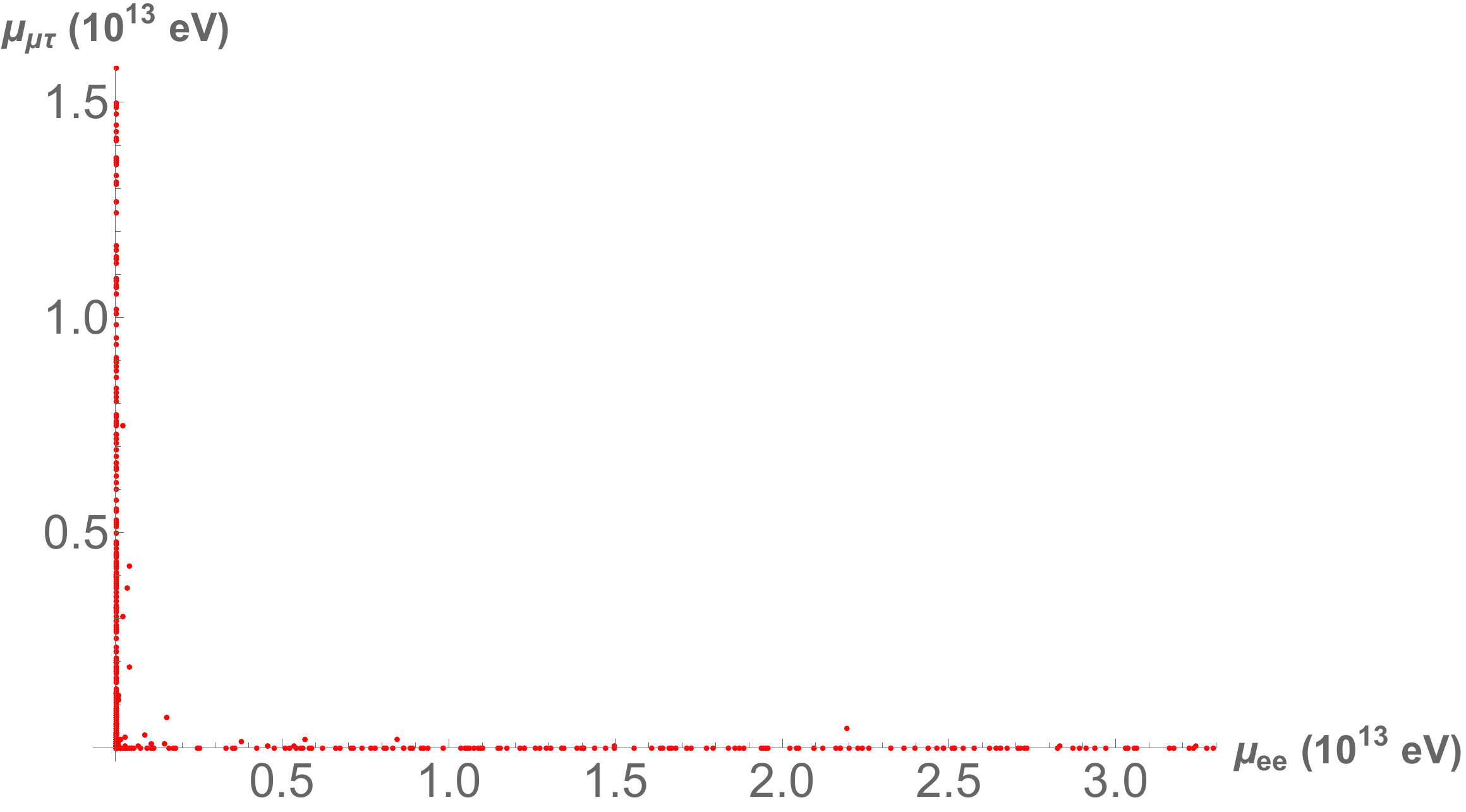}
\caption{The plots show the correlations between free parameters of the model: $\mu_{\mu\tau}$ Vs $\phi$ (left) and $\mu_{\mu\tau}$ Vs $\mu_{ee}$(right) from the parameter scan of the model.}
\label{param1}
\end{figure}
\begin{figure}
\hskip-2.0cm\includegraphics[scale=0.4]{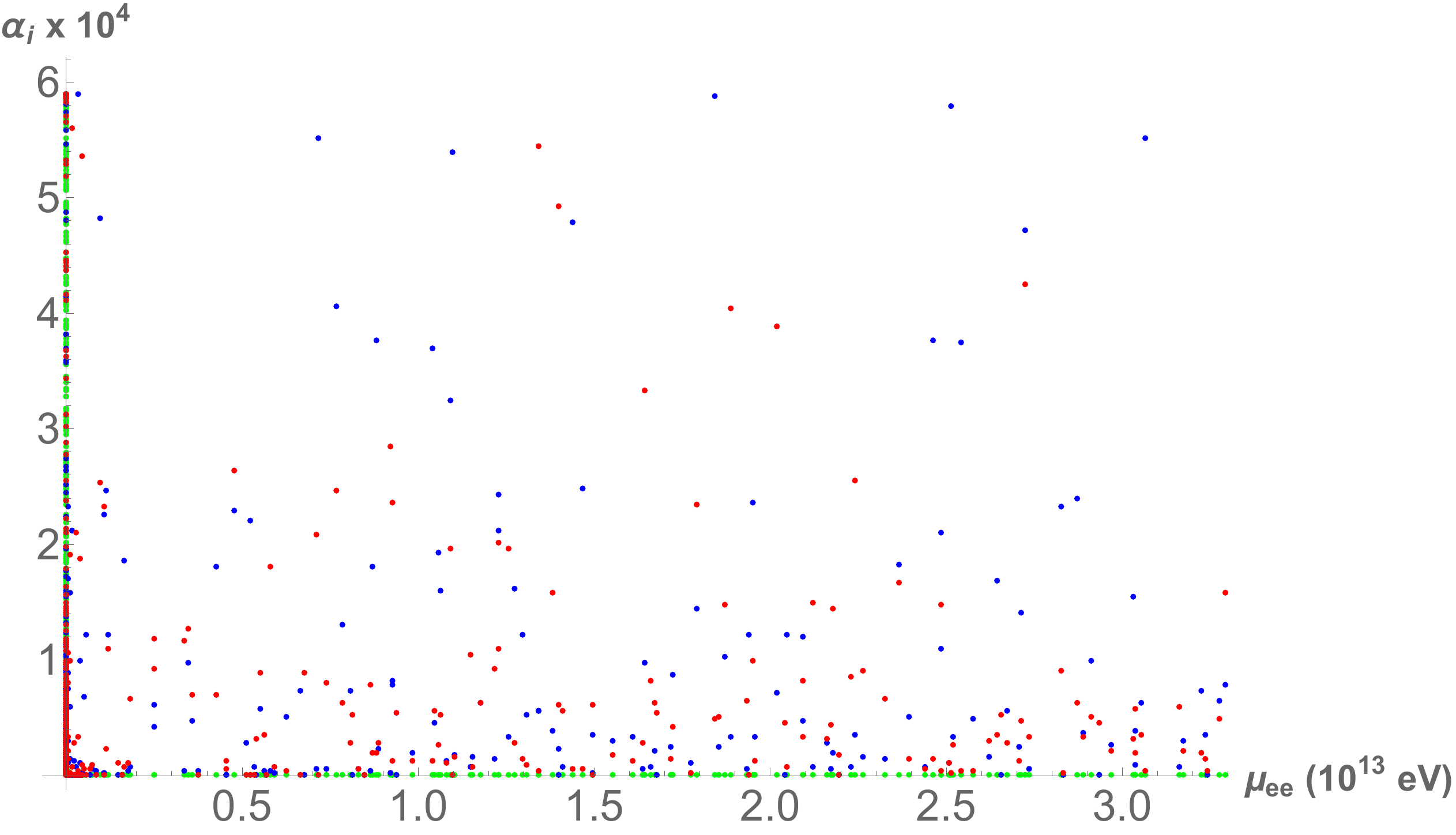}
\includegraphics[scale=0.4]{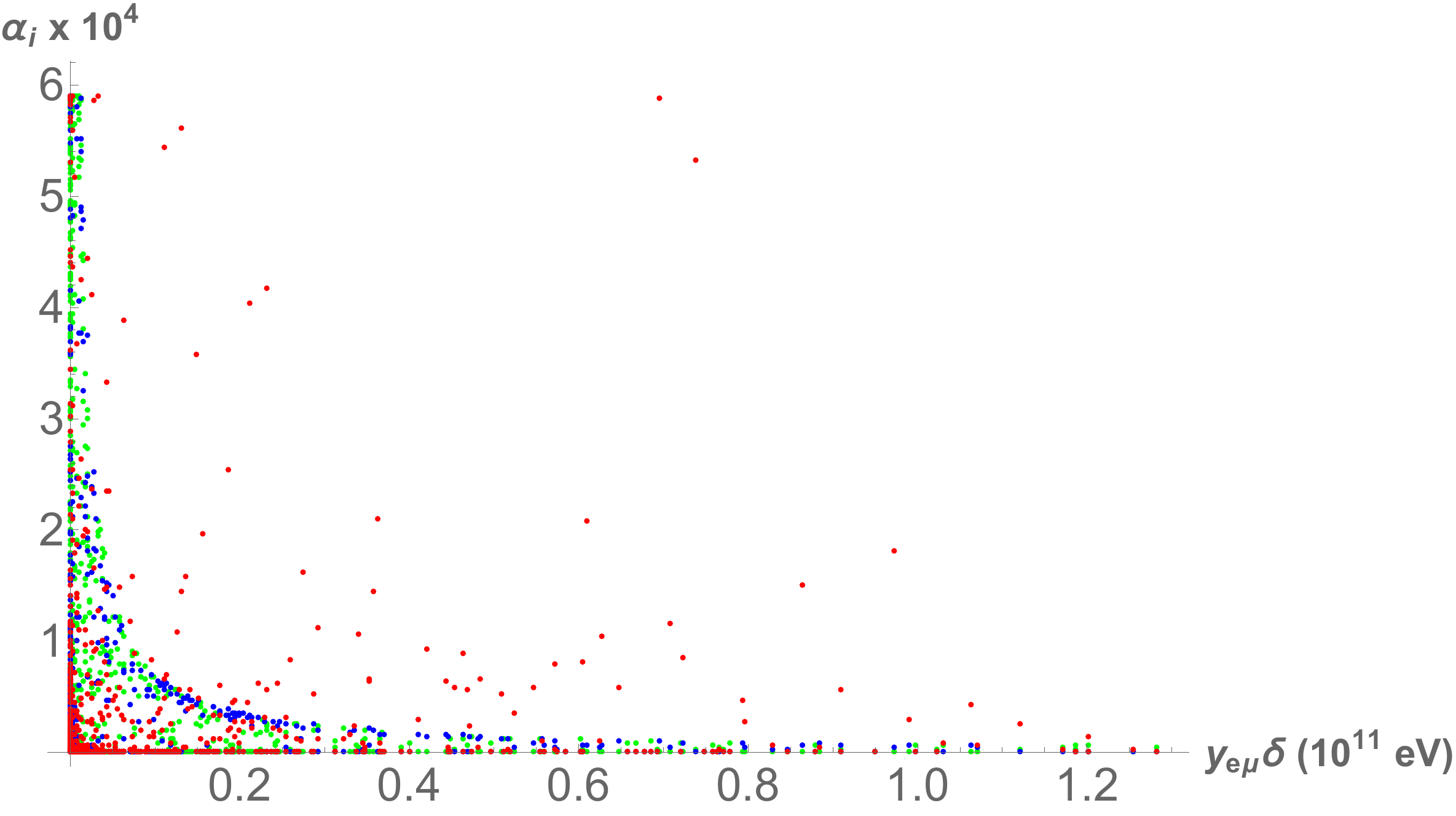}
\caption{The plots show the correlations between some free parameters of the model: $\alpha_{1,2,3}$(green, blue, red) Vs $\mu_{ee}$ (left) and $\alpha_{1,2,3}$(green, blue, red) Vs $y_{e\mu}\delta$(right) from the parameter scan of the model.}
\label{param2}
\end{figure}

In order to check if the matrix in  eq.(\ref{eq1}) can satisfy these constraints, we have done a parameter scan of this model using the following range of values for the model parameters
\begin{itemize}
\begin{item}
We will restrict values for Yukawa couplings, $y_{e\mu}$ and $y_{e\tau}$, to be between $0$ and $\sqrt{4\pi}$. This keeps the couplings in the perturbative regime.
\end{item}
\begin{item}
As discussed in the previous section, we will restrict the VEV of $U(1)_{L_\mu-L_\tau}$ breaking Higgs, $\delta$, to be between $(12-800)$ GeV to resolve the discrepancy in muon (g-2).
\end{item}
\begin{item}
There is no restriction on L-R symmetry breaking scale. In order to have accessible new physics at LHC, we will restrict the VEV of L-R symmetry breaking Higgs, v, to be in the range $(3-10)$ TeV. Assuming values of Yukawa couplings to the typical leptonic values, we can restrict $\alpha_i's$ in the range $(m_{e}/10\text{ TeV},~m_{\tau}/3 \text{ TeV})$=$(5.1\times 10^{-8},~5.9\times 10^{-4})$.
\end{item}
\begin{item}
The value of phase $\phi$ is allowed its full range of $(0-2\pi)$.
\end{item}
\begin{item}
 The Majorana masses, $\mu_{ee}$ and $\mu_{\mu\tau}$, are treated as free parameters which can be as small as needed.
\end{item}
\end{itemize}
A parameter scan of the model shows that it can only accommodate inverted ordering of neutrino masses. A good point in the parameter space that can accommodate the best fit values of all the  oscillation parameters and the muon anomalous magnetic moment is presented in Table.\ref{fit}.
As we will show in the next section, the presence of two texture zeros in the mass matrix does not leave a lot of leeway in the range of neutrino mass matrix entries. This is strongly seen in Fig.\ref{param1} regarding the value of the phase $\phi$ which is restricted to $\pm(0.991-0.999)\pi$ despite being allowed its full range in the parameter scan. All the non-zero entries are of the order of $10^{-2} $eV.  Attempts to keep all the entries of $m_\nu$ in eq.($\ref{eq1}$) $\approx 10^{-2}$ eV results in inverse relationships between $\alpha_1$ and $\mu_{ee}$, $\alpha_{1,2}$ and $y_{e\mu}\delta$ in Fig.\ref{param2} and also that between $\mu_{ee}$ and $\mu_{\mu\tau}$ in Fig.\ref{param1}.
This also results in the Majorana masses $\mu_{\mu\tau}$ and $\mu_{ee}$ taking values in $(10^4-10^{13})$ eV range. Such relationships exist among other free parameters in the model that arise from similar considerations.

\section{Predictions from texture analysis}\label{6}

The $U(1)_{L_\mu-L_\tau}$ gauge symmetry restricts the form of Dirac and Majorana mass matrices leaving the light neutrino mass matrix of the form in eq.(\ref{eq1}). This mass matrix, $m_\nu$, in our model has two zeroes in its $(\mu,\mu)$ and $(\tau,\tau)$ entries. A detailed study of possible neutrino mass matrices with such two-zero structures has been done in \cite{Berger:2000zj, Kageyama:2002zw, Xing:2002ta, Kageyama:2002zw, Xing:2002ap} where seven such patterns that can accommodate the current values of neutrino oscillation parameters are identified. Similar analysis of zero structures in the context of minimal extended type-I seesaw models are done in \cite{Nath:2016mts}. The pattern in our model is usually referred in the literature as the type C pattern. This pattern is also realized in many $U(1)_{L_\mu-L_\tau}$ models where neutrino masses are radiatively generated \cite{Baek:2015mna, Lee:2017ekw} and in \cite{Chen:2017gvf}. In \cite{Baek:2015mna, Lee:2017ekw, Chen:2017gvf}, the type-C pattern and its predictions are discussed in some detail. The two zeroes in $m_\nu$ can give four real relations among the mixing angles, masses, Majorana, and Dirac phases. Using the precisely measured values of the three mixing angles and the two mass-squared differences, it is possible to predict the values of the remaining four parameters: the two Majorana phases ($\alpha_{1,2}$), CP-violating Dirac phase ($\delta_{CP}$), and the absolute values of neutrino masses. In this section, we will show that this peculiar two-zero structure of the light neutrino mass matrix makes this model very predictive.

The light neutrino mass matrix is complex symmetric and can be diagonalized through a bi-unitary transformation, $U^\dagger m_\nu U^*=\text{diag}\{m_1,~m_2,~m_3\}$. Since the charged lepton mass matrix is diagonal, we can identify the diagonalizing matrix of the light neutrino mass matrix $m_\nu$, $U$, to be the PMNS mixing matrix.

%%%%%%%%%%%%%%%%%%%%%%%%%%%%%%%%%%%%%%%%%%

\newpage
\begin{longtable}{ r|   l }
    \hline \hline
\rowcolor[gray]{0.9}\rule{0pt}{4ex} \bf{~~~~~~~~~~~~~~~~Parameters} \hspace{3cm} &\hspace{3cm}\bf{Parameter Value~~~~~~~~~~~~~~}\\
    \hline
\endfirsthead
$g_{\mu\tau}$&$5\times 10^{-4}$\\
$\delta$&12.956 GeV\\
$m_{Z_{\mu\tau}}$&3.2394 MeV\\
$\mu_{ee}$&$20$ MeV\\
$\mu_{\mu\tau}$&$690.347$ MeV\\
$\phi$&$3.1275$ \\
$y_{e\mu}$&$0.0077185$\\
$y_{e\tau}$&$0.0077185$\\
$\alpha_1$&$4.49618\times10^{-5}$\\
$\alpha_2$&$7.09038\times 10^{-6}$\\
$\alpha_3$&$8.46884\times10^{-6}$\\
\end{longtable}
\addtocounter{table}{-2}
\noindent\rule{17cm}{0.4pt}
\begin{equation*}
\text{Using these values,\hspace{1.5in}}M_\nu=\left(
\begin{array}{ccc}
0.0404312 & 0.0318796 & 0.0380774 \\
0.0318796 & 0 & 0.0414535~e^{i~3.1275}\\
0.0380774 &0.0414535~e^{i~3.1275} & 0 \\
\end{array}
\right),
\end{equation*}
\begin{equation*}
\hspace{1in} U_{PMNS}=\left(
\begin{array}{ccc}
 0.809368\, -0.180937 i & 0.376511\, +0.385777 i & -0.146966+0.00101394 i \\
 0.156276\, +0.264408 i & 0.0775003\, -0.568667 i & -0.759101-0.00652307 i \\
 0.37623\, +0.276229 i & 0.176641\, -0.590613 i & 0.634123\, +0.00329828 i \\
\end{array}
\right),\\
\end{equation*}
\begin{equation*}
\hspace{1.6in}U_{PMNS}^\dagger M_\nu U_{PMNS}^*=\left(
\begin{array}{ccc}
m_1=0.0642866 &  &  \\
 &m_2=0.0648573 &  \\
 & &m_3=0.0407998\\
\end{array}
\right).
\end{equation*}
\noindent\rule{17cm}{0.4pt}
\begin{longtable}{ @{\hskip -1in}r|   l@{\hskip 1in}r|l }
$~~~~~~~~~~~~~~~~~~~~~~~~~~~~~~~~~$$\sin^2\theta_{13}$&$0.0216$&$\sum m_{\nu}$&0.169944~eV\\
                                   $\sin^2\theta_{12}$&0.297&$\alpha_2$&$-0.647754\pi$\\
                                   $\sin^2\theta_{23}$&0.589&$\alpha_3$&$+0.489799\pi$\\
                                   $\delta m^2$&$7.37\times 10^{-5}~$eV$^2$&$\phi_1$&0.439873\\
                                   $\Delta m^2$&$-2.505\times 10^{-3}~$eV$^2$&$\phi_2$&2.36362\\
                                   $\delta_{CP}$&$1.31271 \pi$&$\phi_3$&-0.774577\\
                                   $\Delta a_{\mu}$&$29\times 10^{-10}$&$\ev{m_{\beta\beta}}$&0.0404312 eV\\
                                   \\
\caption{\leftskip=-2cm \rightskip=-2.2cm A point in the parameter space that satisfies the experimentally measured best-fit values of mixing angles, mass-squared differences, and the CP-violating Dirac phase given in Table.\ref{data} and muon anomalous magnetic moment given in eq.(\ref{MAM}) is tabulated above. The values of Majorana phases $\alpha_{2,3}$, neutrino masses $m_i$, the sum of neutrino masses $\sum m$, and the effective Majorana mass of neutrinoless double beta decay $\ev{m_{\beta\beta}}$ are also given. The unphysical phases $\phi_{1,2,3}$ that arise during diagonalization are also tabulated for completeness.}
\label{fit}
\end{longtable}

\newpage

\hskip-4cm\begin{figure}

\begin{minipage}{.5\linewidth}
\centering
\subfloat[]{\label{main:a}\includegraphics[scale=.25]{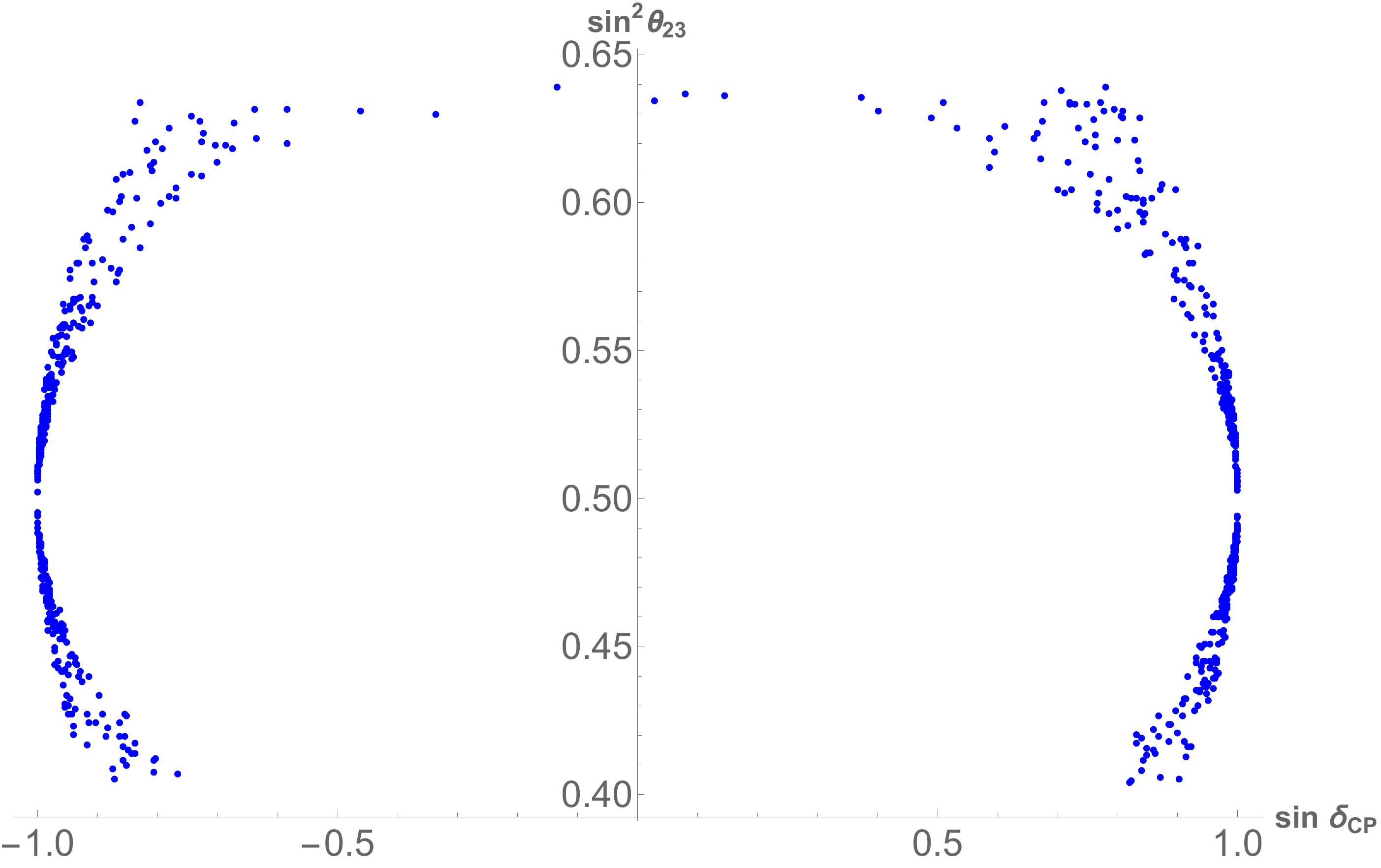}}
\end{minipage}%
\begin{minipage}{.5\linewidth}
\centering
\subfloat[]{\label{main:b}\includegraphics[scale=.25]{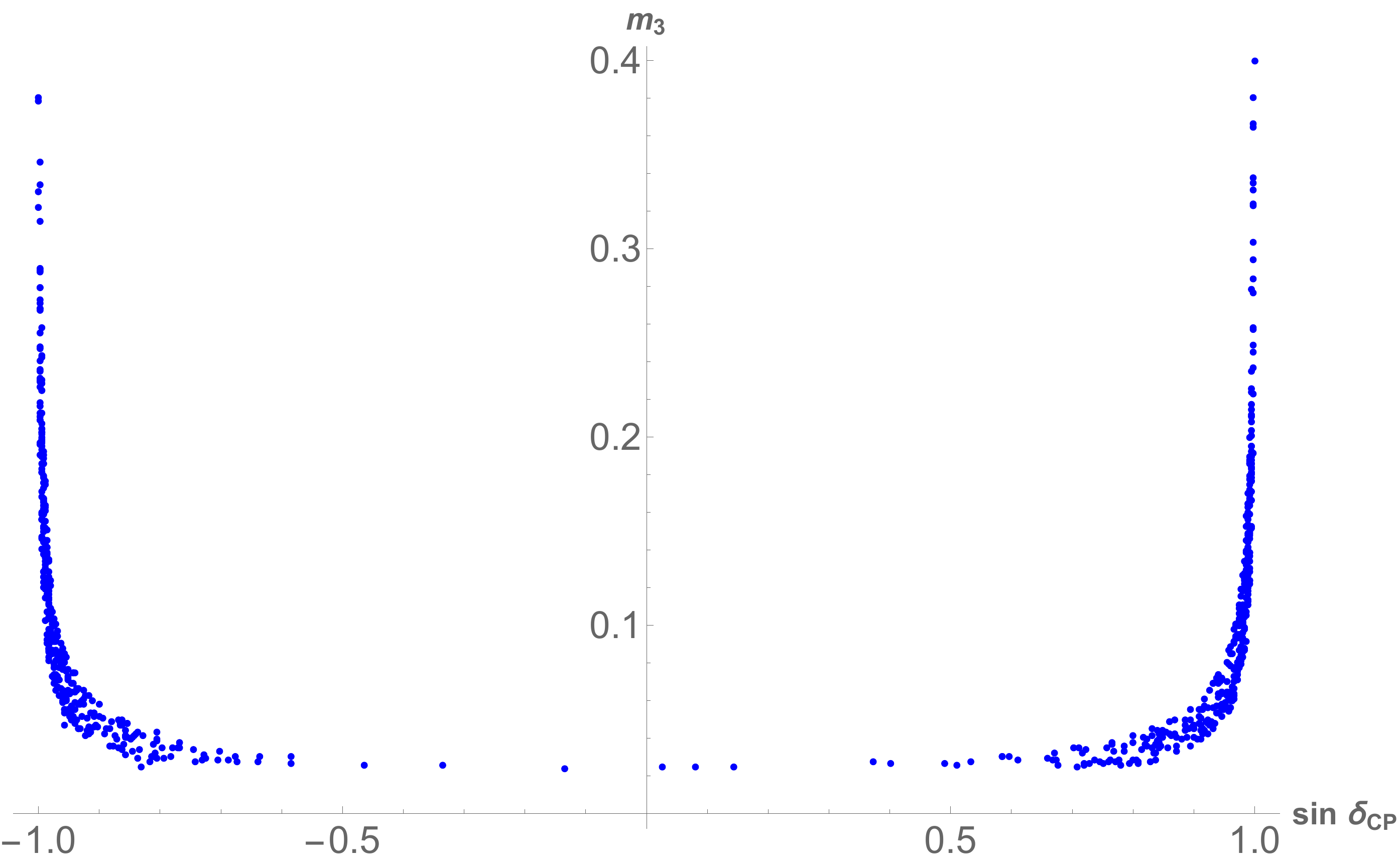}}
\end{minipage}\par\medskip
\centering
\subfloat[]{\label{main:c}\includegraphics[scale=.3]{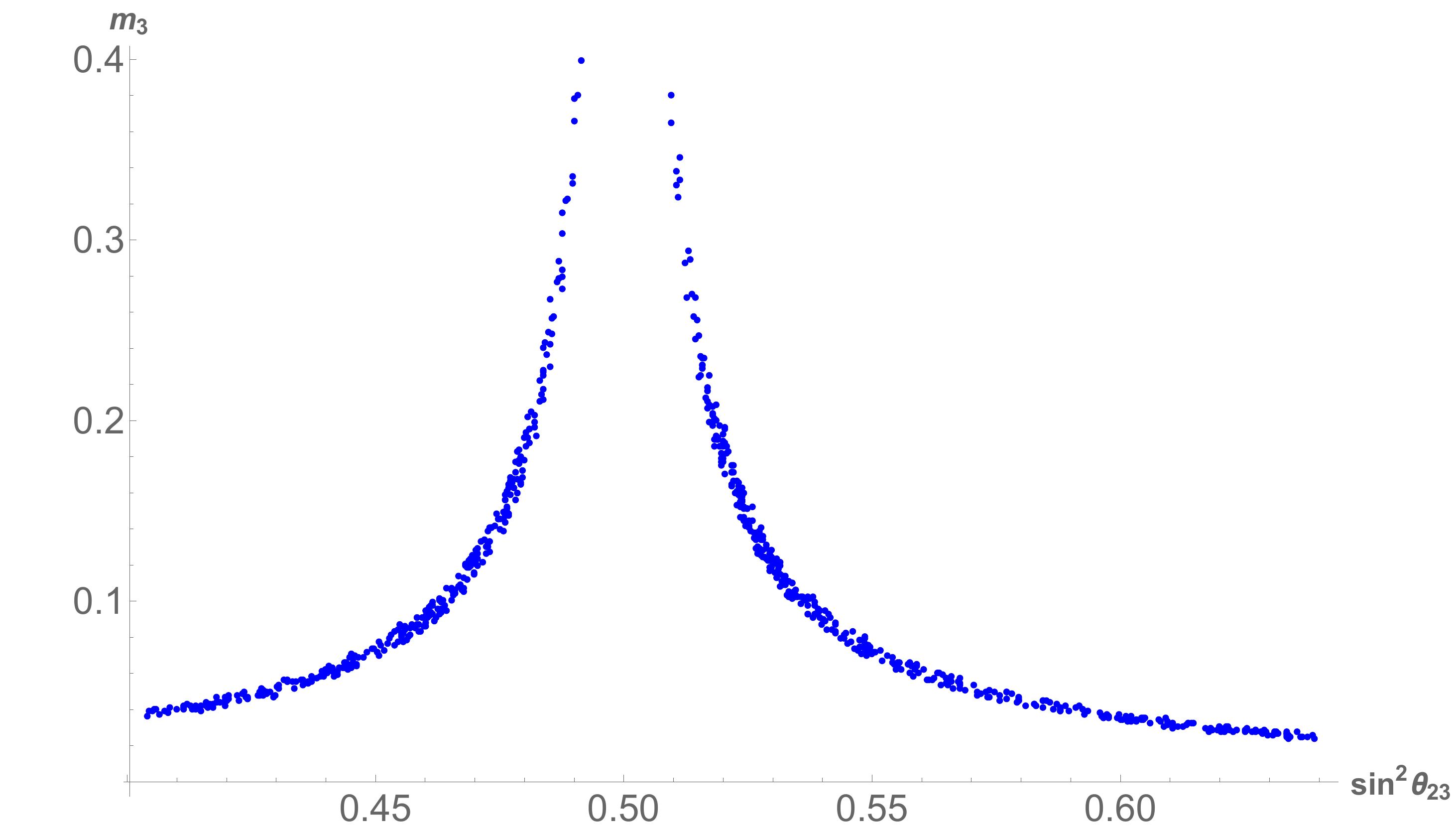}}

\caption{The plots show correlations between various oscillation parameters: (a) $\sin^2\theta_{23}~\text{Vs}~\sin \delta_{CP}$, (b) $m_3~\text{Vs}~\sin \delta_{CP}$, and (c) $m_3~\text{Vs}~\sin^2\theta_{23}$ from the parameter scan of the model.}
\label{correl}
\end{figure}

Using the standard parametrization of the $U_{PMNS}$ matrix:
\begin{equation}\label{param}
\left(\begin{array}{c c c}
e^{i\phi_1}&~&~\\
&e^{i\phi_2}&\\
&&e^{i\phi_3}\end{array}\right)\left(\begin{array}{c c c} 
c_{12}c_{13}~ & ~s_{12}c_{13} ~&~ s_{13}e^{-i \delta}\\
-s_{12}c_{23}-c_{12}s_{23}s_{13}e^{i\delta}~&~c_{12}c_{23}-s_{12}s_{23}s_{13}e^{i\delta}~&~s_{23}c_{13}\\
s_{12}s_{23}-c_{12}c_{23}s_{13}e^{i\delta}~&~-c_{12}s_{23}-s_{12}c_{23}s_{13}e^{i\delta}~&~c_{23}c_{13}
\end{array}
\right)\left(\begin{array}{c c c}
1&~&~\\
&e^{i\frac{\alpha_2}{2}}&\\
&&e^{i\frac{\alpha_3}{2}}\end{array}\right),
\end{equation}the diagonalization conditions lead to two complex equations corresponding to the two zeroes in the neutrino mass matrix
\begin{eqnarray}\label{CE}\notag
m_1 V_{\mu 1}^2  + m_2 e^{i\alpha_2} V^2_{\mu 2}+m_3 V^2_{\mu 3} e^{i\alpha_3}&=&0,\\
m_1 V_{\tau 1}^2 + m_2 e^{i\alpha_2} V^2_{\tau 2}+m_3 V^2_{\tau 3} e^{i\alpha_3}&=&0,
\end{eqnarray}
where $V$ is the CKM part of the $U_{PMNS}$ matrix. The full diagonalizing matrix $U$ has Majorana phases $\alpha_{2,3}$ and unphysical phases $\phi_{1,2,3}$ such that $U=\text{diag}\{\phi_1,~\phi_2,~\phi_3\}.V.\text{diag}\{1,~\alpha_2/2,~\alpha_3/2\}$. The unphysical phases are absorbed into the left-handed charged lepton fields. We note that other studies of two-zero textures use a different parameterization for the $U_{PMNS}$ matrix with regards to the Majorana phases. The use of parameterization in eq.(\ref{param}) makes it easier to compare our results with the results of the type-I seesaw model in \cite{Asai:2017ryy}. The two complex equations in eq.(\ref{CE}) lead to four real relations which relate the three masses, three mixing angles, the CP-violating Dirac phase and the two Majorana phases. The unphysical phases $\phi_{1,2,3}$ which arise during the diagonalization can also be uniquely determined as a function of mixing angles and mass-squared differences by demanding that $(e,e),(e,\mu),$ and $(e,\tau)$ entries of mass matrix $m_\nu$ are real. Since all the neutrino oscillation parameters including the unphysical phases can be uniquely determined as a function of mass-squared differences and mixing angles, the entries of $m_\nu$ are tightly constrained by the current experimental data in its $3\sigma$ range. This can be particularly seen in the case of the free model parameter $\phi$ which is confined to the regions near $\pm \pi$ despite being allowed its full range during the parameter scan as seen in Fig.\ref{param2}.

The following analysis does not depend on the values of the parameters of the model such as the Yukawa couplings or the breaking scale of $U(1)_{L_\mu-\tau}$. The results in this section are applicable to any model with type-C two-zero texture pattern.  

 The four relations in eq.(\ref{CE}) can be re-phrased as
\begin{equation}\label{eqa}
\frac{m_2}{m_1}e^{i\alpha_2}=R_{2},~~\frac{m_3}{m_1}e^{i \alpha_3}=R_{3}.
\end{equation}
where $R_{1}$ and $R_{2}$ are functions of mixing angles and the CP-violating Dirac phase
\begin{eqnarray}\label{massratios}\notag
R_2&=&1+\frac{\sec ^2\left(\theta _{12}\right)}{-1+e^{i \delta } \sin \left(\theta _{13}\right) \tan \left(\theta _{12}\right) \tan \left(2 \theta _{23}\right)},\\
R_3&=&\frac{e^{i \delta } \left(e^{2 i \delta } \tan ^2\theta _{13}-e^{i \delta } \cos 2 \theta _{12} \tan \theta _{13} \cot 2 \theta _{23} \csc \theta _{12} \sec \theta _{12} \sec \theta _{13}+\sec ^2\theta _{13}\right)}{\cot \theta _{12} \cot 2 \theta _{23} \csc \theta _{13} -e^{i \delta }}.
\end{eqnarray}

 \begin{figure}
\hskip-1.25cm\includegraphics[scale=0.28]{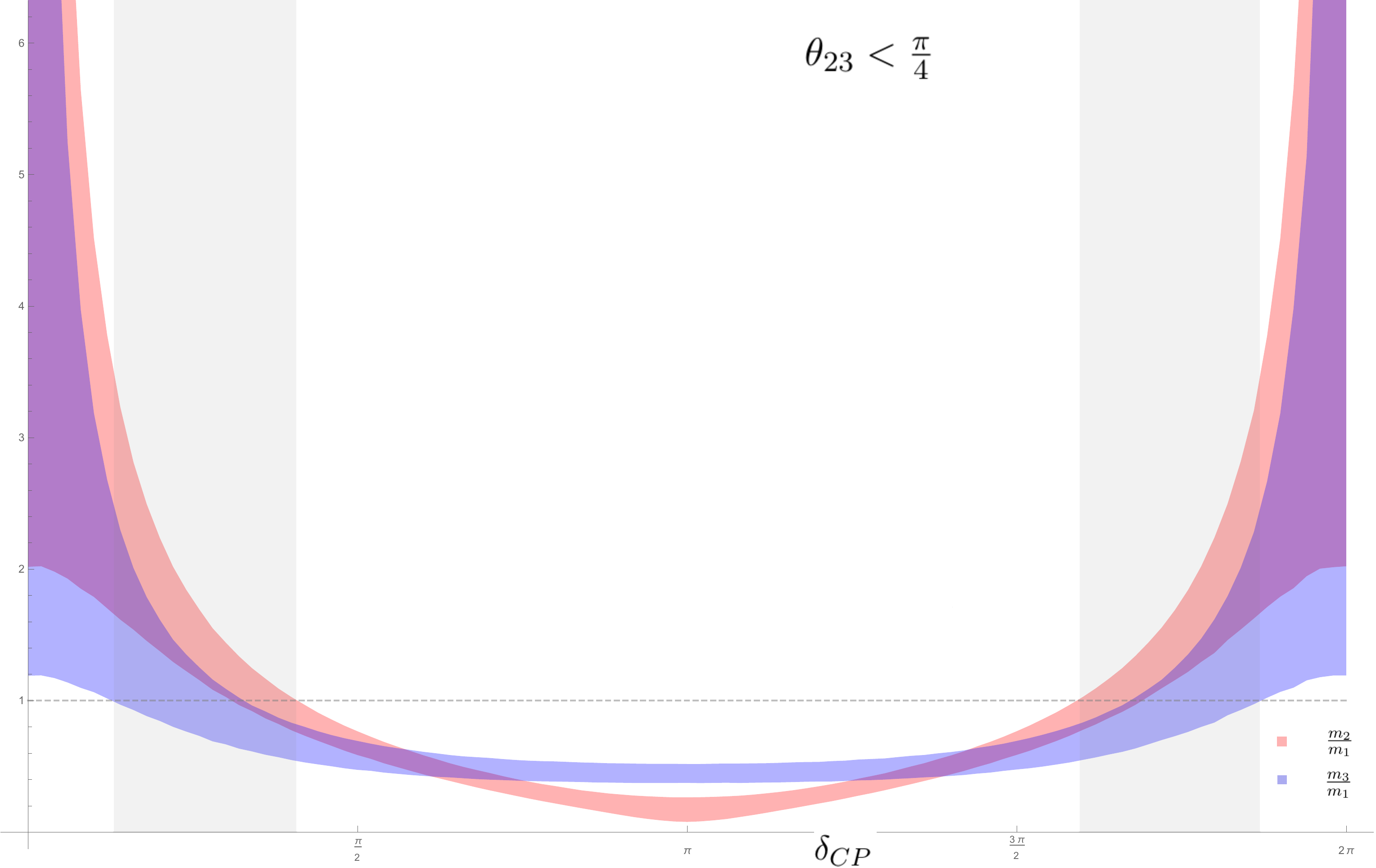}
\includegraphics[scale=0.31]{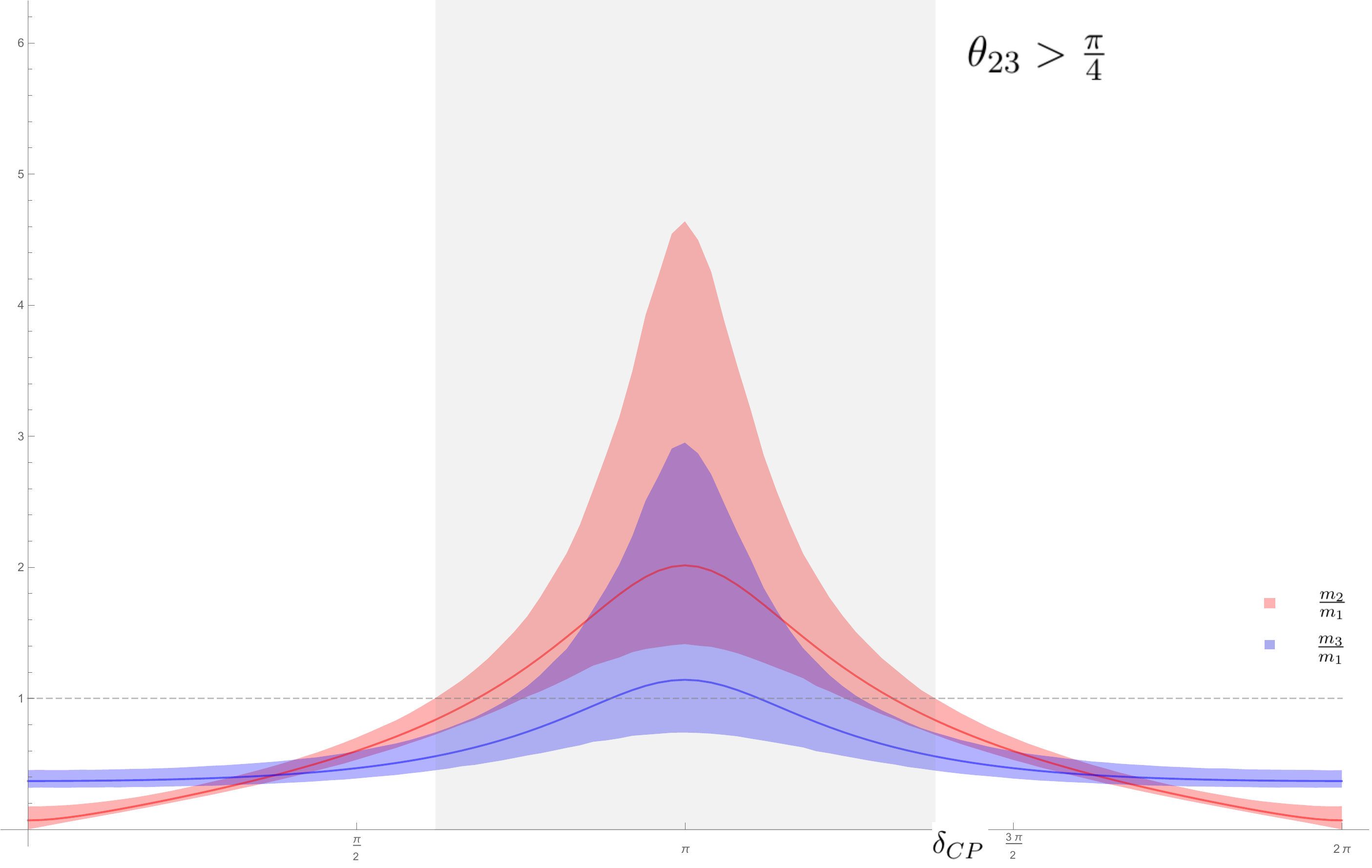}
\caption{The mass ratios $m_2/m_1$(red) and $m_3/m_1$(blue) as a function of $\delta_{CP}$ for values of $\theta_{23}<\pi/4$(left) and $\theta_{23}>\pi/4$(right). The red and blue bands(lines) come from the $1\sigma$ range(central value) in the measurement of mixing angles and mass-squared differences. The gray bands show the regions where inverted ordering is satisfied. The dashed gray line corresponds to $m_{2,3}/m_1=1.$ }
\label{massratiofig}
\end{figure}

 The mass ratios, $m_{2,3}/m_1$, are given by the absolute values of $R_{2,3}$ in eq.(\ref{massratios}). In Fig.\ref{massratiofig}, these ratios are plotted as a function of $\delta_{CP}$ while keeping other oscillation parameters in their $1\sigma$ range as given in the Table.\ref{data}. From Fig.\ref{massratiofig}, we can see that only inverted ordering (IO) is allowed in our model. The value of mass ratios is around unity in this region hinting at a quasi-degenerate hierarchy of light neutrino masses with $m_3\lesssim m_1\lesssim m_2$. Unlike the case of normal ordering, there is still an ambiguity surrounding the octant of $\theta_{23}$ in the case of inverted ordering. We have made separate plots of mass ratios for both choices of octants. It can be seen that both octants accommodate complementary values for $\delta_{CP}$ allowing octant sensitivity in our model. In fact, the current measured value of $\delta_{CP}$ at $1\sigma$ slightly favors $\theta_{23}>\pi/4$. This is complementary to the case of normal ordering where the experiments favor $\theta_{23}<\pi/4.$ A precise measurement of $\delta_{CP}$ will result in a conclusive prediction of the octant of $\theta_{23}$ in our model.
 
In \cite{Capozzi:2017ipn}, the neutrino oscillation parameters are reported in terms of mixing angles, $\theta_{ij}$, CP-violating $\delta_{CP}$, and the mass-squared differences $\delta m^2$ and $\Delta m^2$ defined as $m_2^2-m_1^2$ and $m_3^2-(m_2^2+m_1^2)/2$ respectively. The whole parameter space shown in Fig.\ref{massratiofig} is not accessible as additional constraints from mass-squared differences $\delta m^2$ and $\Delta m^2$ narrows it down further.
 The values of $\delta m^2$ and $\Delta m^2$ provides additional relations in terms of mass ratios
\begin{eqnarray}\label{Dirac}\notag
\delta m^2&=&m_1^2(\frac{1}{|R_2(\delta)|^2}-1),\\
\Delta m^2 + \frac{\delta m^2}{2}&=&m_1^2(\frac{1}{|R_3(\delta)|^2}-1),
\end{eqnarray}
which can be  solved to find the values of the light neutrino masses $m_i$ and the $\delta_{CP}$.

The complex functions $R_{2,3}$ have some interesting symmetry properties with respect to its arguments $\theta_{23}$ and $\delta_{CP}$. Since $\delta$ is the only phase appearing in $R_{2,3}$, the absolute value of complex functions $R_{2,3}$ has the property $|R_{2,3}(\delta)|=|R_{2,3}(2\pi-\delta)|$. As seen in Fig.\ref{massratiofig}, this makes the mass ratios $m_{2,3}/m_1$ symmetric about $\delta=\pi$. However, the arguments of the complex functions, the Majorana phases $\alpha_{2,3}$, get inverted about $\delta=\pi$, i.e., $\alpha_{2,3}(2\pi-\delta)= -\alpha_{2,3}(\delta)$. Since $\delta_{CP}$ enters eqs.(\ref{Dirac}) through $|R_{2,3}(\delta)|$, any set of choice for parameters $\theta_{ij}$, $\delta m^2$, and $\Delta m^2$ will accommodate two solutions, $\delta$ and $2\pi-\delta$, and their corresponding Majorana phases $\pm\alpha_{2,3}$. Another subtle property is the invariance of $|R_{2,3}|$ under the simultaneous transformations $\tan(2\theta_{23})\rightarrow-\tan(2\theta_{23})$ and $e^{i\delta}\rightarrow-e^{-i\delta}$. 
\begin{table}
\hskip-1.0cm\begin{tabular}{cc c c c}
\hline\hline
Parameter&best-fit&$1\sigma$&$2\sigma$&$3\sigma$\\
\hline\\
$\delta m^2/10^{-5}$eV$^2$&7.37&7.21-7.54&7.07-7.73&6.93-7.96\\
\\
$\sin^2\theta_{12}/10^{-1}$&2.97&2.81-3.14&2.65-3.34&2.50-3.54\\
\\
|$\Delta m^2$|$/10^{-3}$ eV$^2$& 2.505 & 2.473 - 2.539 & 2.430 - 2.582 & 2.390 - 2.624 \\
\\
$\sin^2 \theta_{13}/10^{-2}$
                           & 2.16 & 2.07 - 2.24 & 1.98 - 2.33 & 1.90 - 2.42 \\
\\
$\sin^2 \theta_{23}/10^{-1}$ 
                             & 5.89 & 4.17 - 4.48 $\oplus$ 5.67 - 6.05 & 3.99 - 4.83 $\oplus$ 5.33 - 6.21 &  3.84 - 6.36 \\
\\
$\delta_{CP}/\pi$ 
            & 1.31 & 1.12 - 1.62 & 0.92 - 1.88  &  0 - 0.15 $\oplus$ 0.69 - 2  \\
             
          \hline\hline\\
 %\hline%-----------------------------------
\end{tabular}
\caption{The values of neutrino oscillation parameters in their best-fit, $1\sigma,~2\sigma,$ and $3\sigma$ ranges from the global $3\nu$ oscillation analysis with inverted ordering for neutrino masses \protect\cite{Capozzi:2017ipn}.}
\label{data}
\end{table}

 \subsection{Results}

\begin{figure}
\hskip-1cm\includegraphics[scale=0.45]{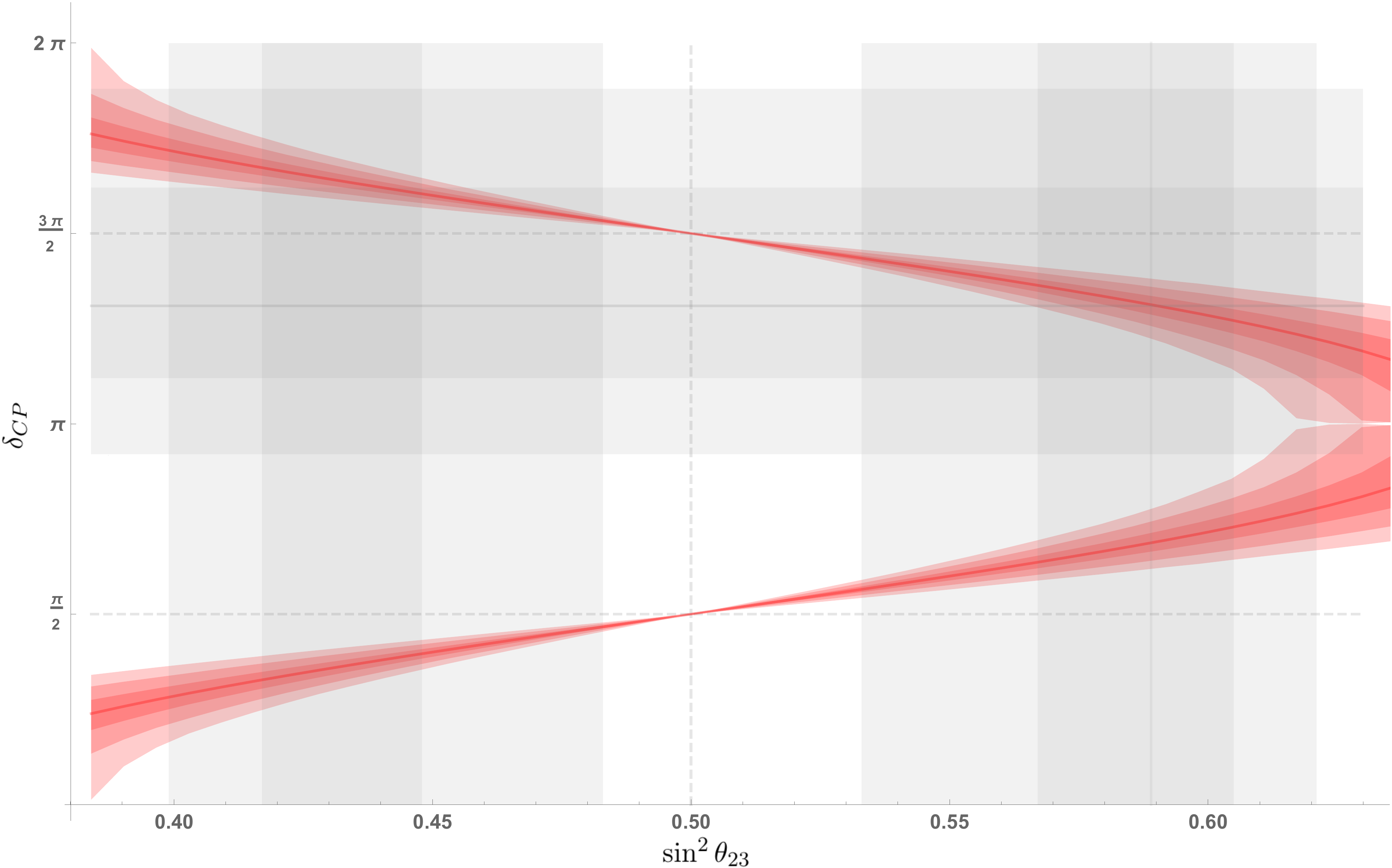}
\caption{The plot shows the correlation between CP-violating Dirac phase $\delta_{CP}$ and $\sin^2\theta_{23}$ in their $3\sigma$ ranges in the type-C two-zero texture pattern. The red bands(line) show(s) the uncertainty (best-fit) coming from $1,~2,~3\sigma$ errors(central values) of other mixing angles and mass-squared differences. The horizontal and vertical gray bands(line) show(s) the $1,~2\sigma$ ranges(central value) of $\delta_{CP}$ and $\sin^2\theta_{23}$ respectively. The dashed gray lines highlight some interesting features of the plot.}
\label{dcpvst23}
\end{figure}

The plot of calculated $\delta_{CP}$ as a function of $\theta_{23}$ in its $3\sigma$ range is given in Fig.\ref{dcpvst23}. As expected from the symmetries of $R_{2,3}$, there are two solutions $\delta$ and $2\pi-\delta$ for any choice of input parameters. The red bands show the uncertainty due to the error in the measured values of input parameters given in Table.\ref{data}. For $\theta_{23}=\pi/4,$ $\delta_{CP}$ is predicted to be $\pi/2,3\pi/2$ irrespective of the values of other input parameters. The calculated best-fit ($1,~2,3\sigma$) value (range) for $\delta_{CP}$ is given in Table.\ref{predictions}. It is remarkable to notice that one of the best-fit predictions for $\delta_{CP}$, $1.31\pi$, agrees with the experimentally measured best-fit value. As seen from Fig.\ref{dcpvst23}, the current measured  best-fit and $1\sigma$ range of $\delta_{CP}$ mostly hint towards a value of $\theta_{23}>\pi/4$. 

\begin{figure}
\hskip-2.0cm\includegraphics[scale=0.48]{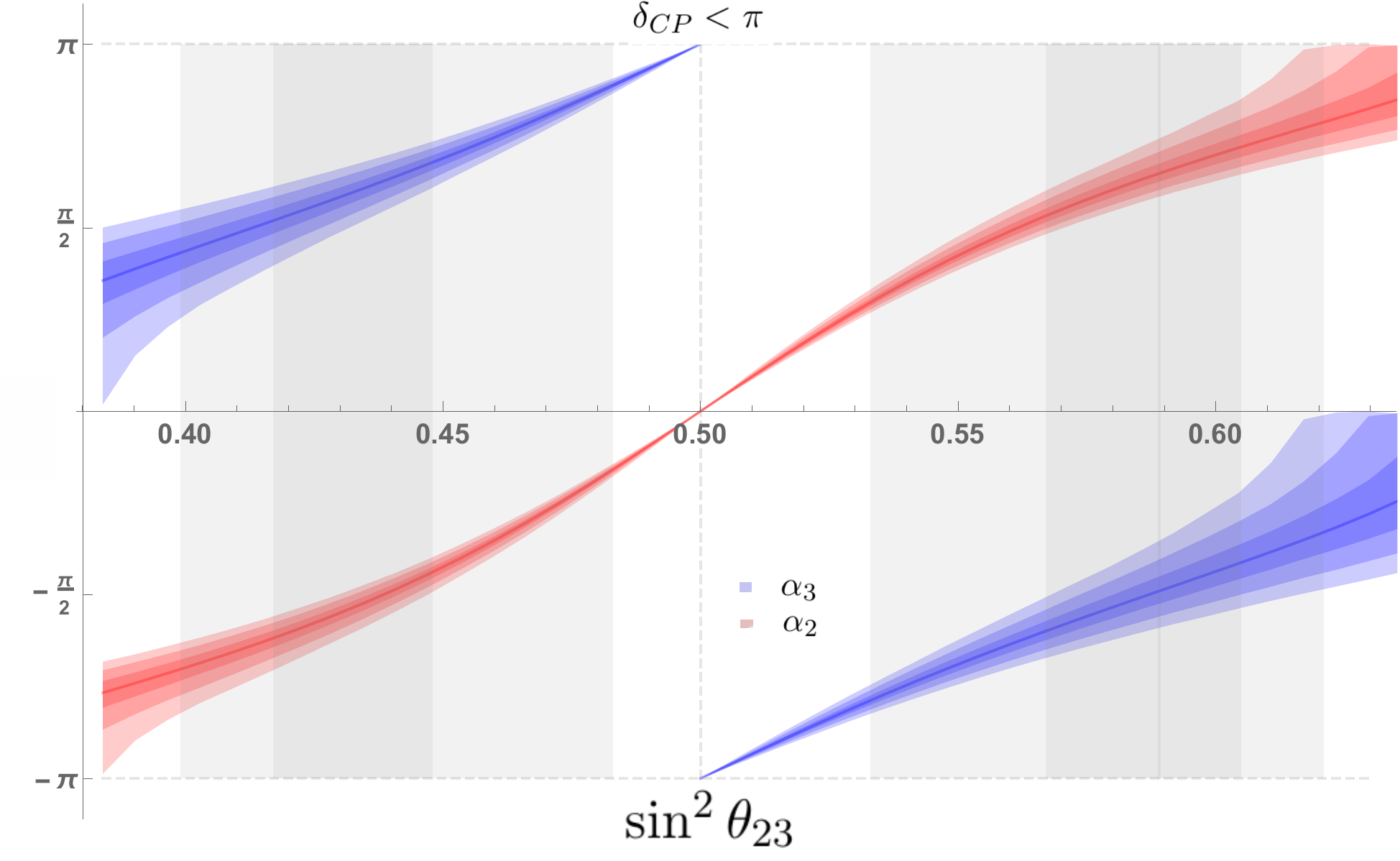}
\includegraphics[scale=0.48]{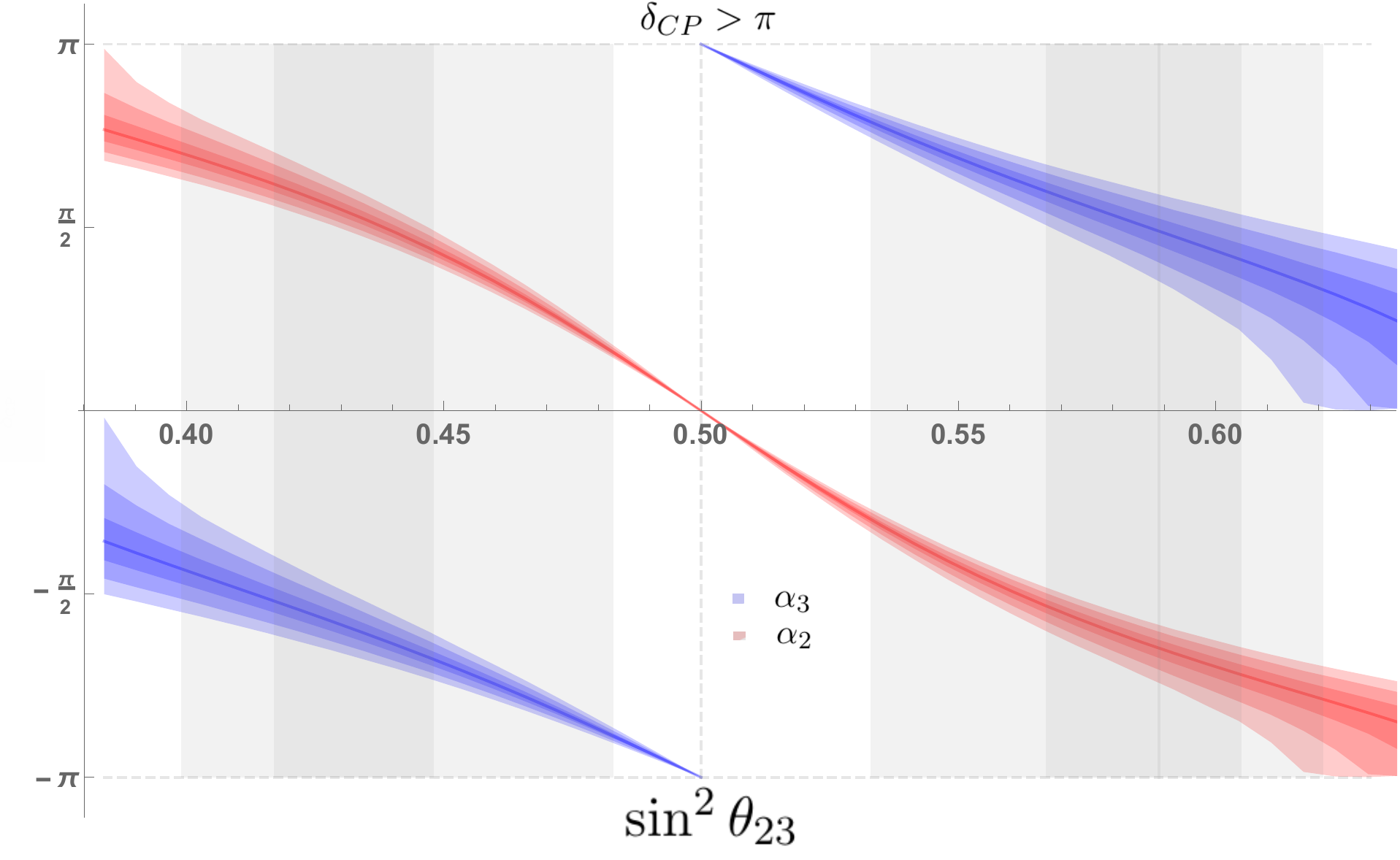}
\caption{The plot shows the correlation between Majorana phases $\alpha_{2,3}$ (red, blue) and $\sin^2\theta_{23}$ in $3\sigma$ range of $\sin^2 \theta_{23}$ in the type-C two-zero texture pattern. The red and blue bands(lines) show the uncertainty(best-fit) coming from $1,~2,3\sigma$ errors(central values) of other mixing angles and mass-squared differences. The gray bands(line) show(s) the $1,~2\sigma$ ranges(central value) of $\sin^2\theta_{23}$. The dashed gray lines highlight some interesting features of the plot.}
\label{al23vst23}
\end{figure}

 For any choice of parameters, we get two solutions for the Majorana phases $\pm \alpha_{2,3}$ corresponding to the two solutions for $\delta_{CP}$. Separate plots of the Majorana phases as a function of $\theta_{23}$ for $\delta_{CP}<\pi$ and $\delta_{CP}>\pi$ are shown in Fig.\ref{al23vst23}. The blue and the red bands denote the uncertainty coming form the $n\sigma$ errors in the measured value of mixing angles and mass-squared differences. The best-fit ($1\sigma,~2\sigma,~3\sigma$) value(ranges) of $\alpha_{2}/\pi$ and $\alpha_3/\pi$ are $\pm 0.65(\pm(0.4-0.75),~\pm (0.16-0.9),~(-1,1))$ and $\pm 0.49(\pm(0.35-0.7),~\pm (0.1-0.9),~(-1,1))$ respectively.  
 \begin{figure}
\centering
\includegraphics[scale=0.4]{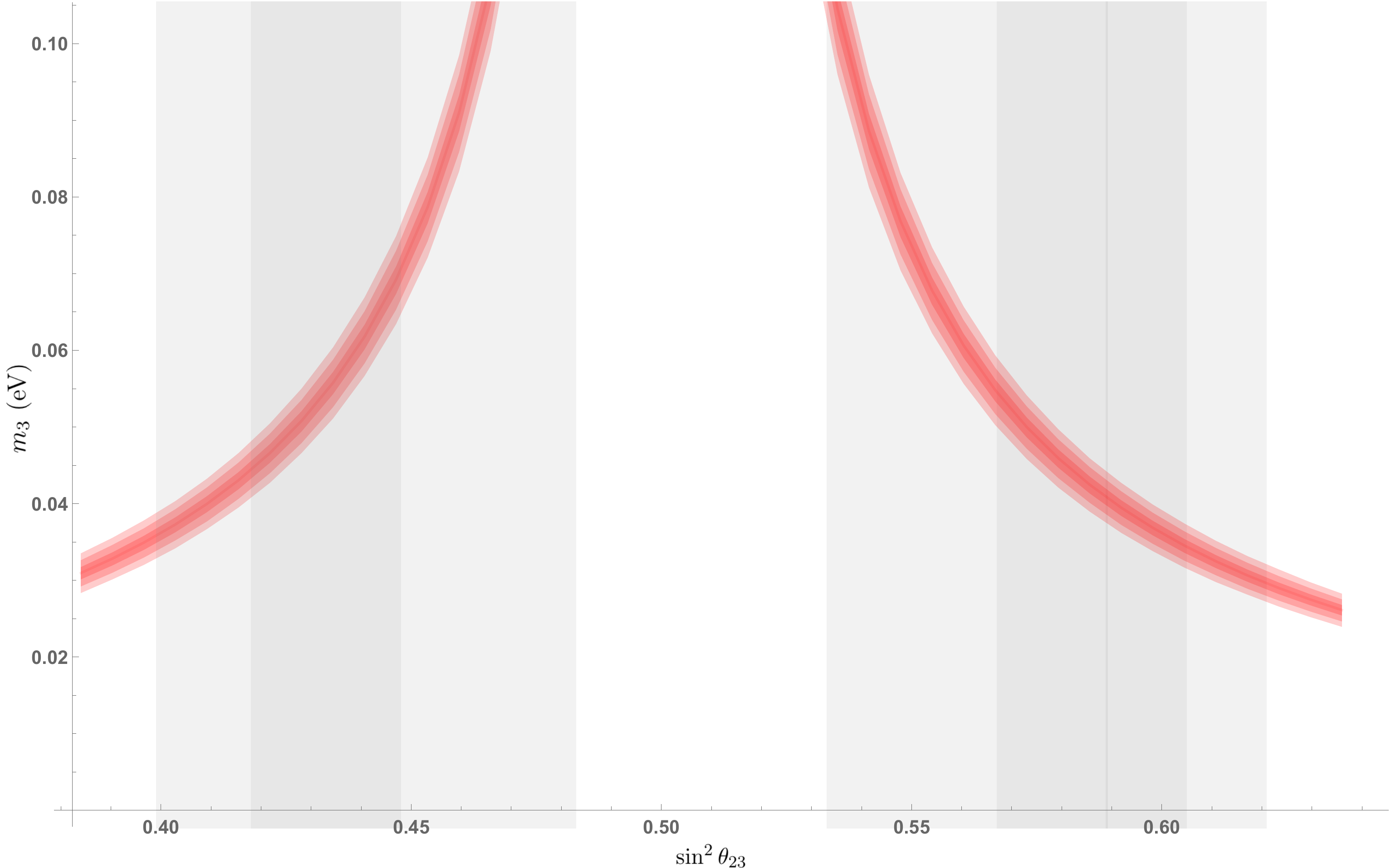}
\caption{The plot shows the correlation between the mass of lightest neutrino $m_{3}$ and $\sin^2\theta_{23}$ in the $3\sigma$ ranges of $\sin^2\theta_{23}$ in the type-C two-zero texture pattern. The red bands(line) show(s) the uncertainty(best-fit) coming from $1,~2,3\sigma$ errors(central values) of other mixing angles and mass-squared differences. The gray bands(line) show(s) the $1,~2\sigma$ ranges(central value) of $\sin^2\theta_{23}$. The dashed gray lines highlight some interesting features of the plot.}
\label{m3vst23}
\end{figure}

\begin{figure}
\centering
\includegraphics[scale=0.55]{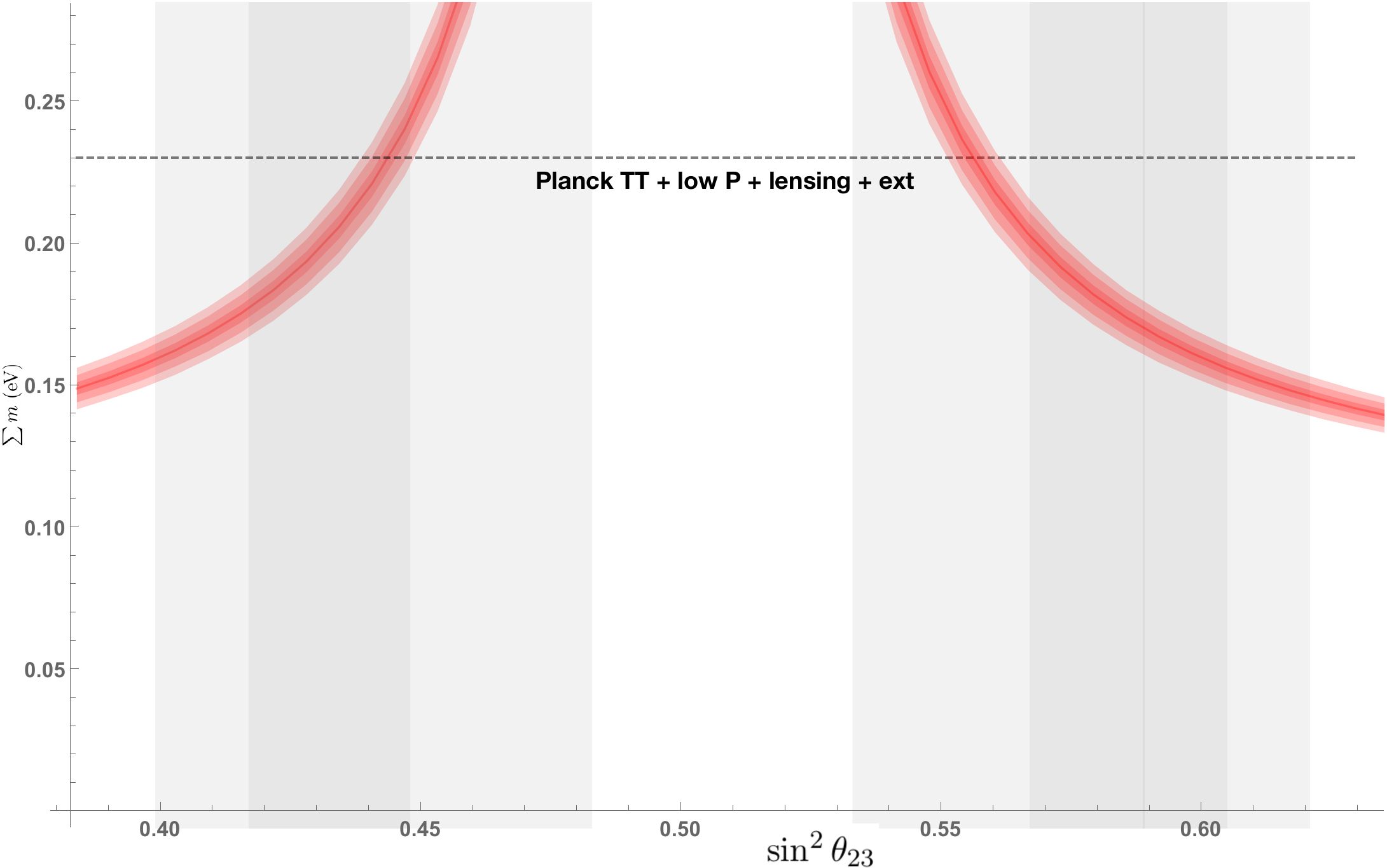}
\caption{The prediction for the sum of neutrino masses, $\sum m$, as a function of $\sin^2\theta_{23}$. The red bands(line) show(s) the uncertainty(best-fit) due to $1,~2,3\sigma$ errors(central values) of other mixing angles and mass-squared differences. The gray bands(line) show(s) the $1,~2\sigma$ ranges(central value) of $\sin^2\theta_{23}$. The dashed black line shows the current cosmological bound on $\sum m$ from Planck TT, low P, and the lensing experiments.}
\label{summvst23}
\end{figure}

By solving eq.(\ref{Dirac}) and using the definitions of $\delta m^2$ and $\Delta m^2,$ the mass spectrum of light neutrinos can be determined. The mass of lightest neutrino in the inverted order, $m_3$, is plotted as a function of $\theta_{23}$ in Fig.\ref{m3vst23}. As expected from symmetries, the plot is symmetric about $\theta_{23}=\pi/4$. The red bands show the uncertainty coming from the error in the measured value of mixing angles and mass-squared differences. The best-fit ($1\sigma,~2\sigma,~3\sigma$) value(ranges) of $m_3$ is calculated as $0.04((0.03-0.07),~(0.027-0.24),~(0.026-\infty))$. As seen from eqs.\ref{massratios}, $m_2/m_1=1$ at $\theta_{23}=\pi/4$. In order to obtain a finite mass splitting, the mass blows up at $\theta_{23}=\pi/4$.
 The model, therefore, predicts relatively large values for neutrino masses. The plot of the sum of the light neutrino masses, $\sum m$, as a function of $\theta_{23}$ along with the $n\sigma$ uncertainty bands from the measurement of other mixing angles and mass-squared differences, is given in Fig.\ref{summvst23}. 
A strong constraint on the absolute scale of neutrino masses comes from the cosmological constraint on the sum of light neutrino masses. Results from Planck experiment\cite{Ade:2015xua} reports an upper bound on the sum of neutrino masses of $0.23$ eV. This upper bound is shown in Fig.\ref{summvst23} as a black dashed line which eliminates a large region of parameter space in $3\sigma$ range of $\theta_{23}$ but leaves most of the $1\sigma$ region in both octants largely untouched. However, recent studies of Planck data report more stringent upper bounds of 0.18 eV \cite{Giusarma:2016phn} and 0.15 eV \cite{Vagnozzi:2017ovm} with analysis of polarization data bringing it as low as 0.12 eV \cite{Vagnozzi:2017ovm}. Future cosmological data will be one of the first challenges facing this model The best-fit ($1\sigma,~2\sigma,~3\sigma$) value(ranges) of $\sum m$ is given by  $0.169((0.153-0.248),~(0.141-0.71),~(0.133-\infty))$.

\begin{figure}
\centering
\includegraphics[scale=0.55]{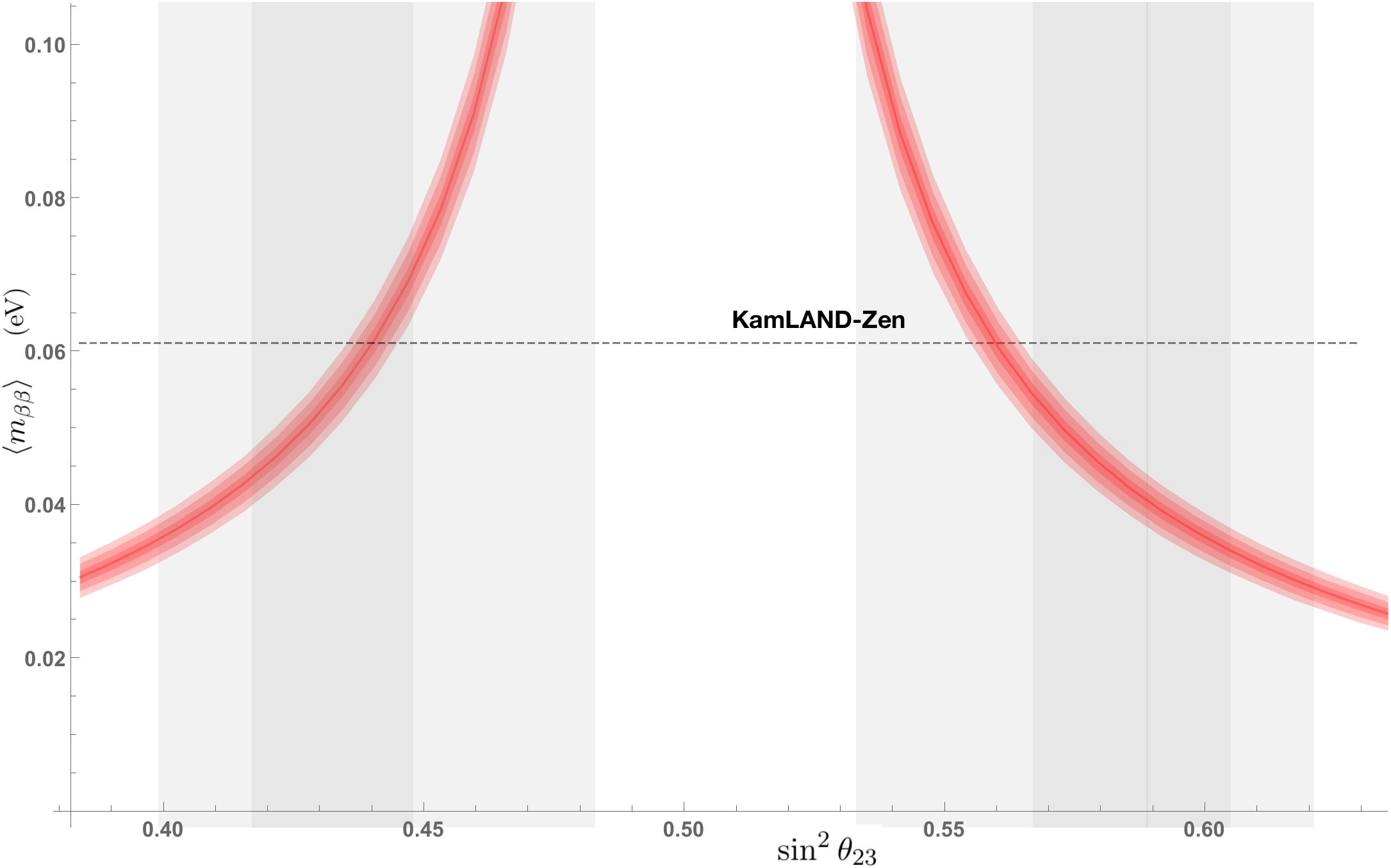}
\caption{The prediction for the effective Majorana mass of neutrinoless double beta decay $\ev{m_{\beta \beta}}$ as a function of $\sin^2\theta_{23}$. The red bands(line) show(s) the uncertainty(best-fit) coming from $1,~2,~3\sigma$ errors (central values) of other mixing angles and mass-squared differences. The gray bands(line) show(s) the $1,~2\sigma$ ranges(central value) of $\sin^2\theta_{23}$. The dashed black line shows the strongest bound on $\ev{m_{\beta \beta}}$ from KamLAND-Zen\protect\cite{KamLAND-Zen:2016pfg}.}
\label{mbbvst23}
\end{figure}

 Another strong constraint on models with large light neutrino masses comes from the search of neutrinoless double beta decay. The effective Majorana neutrino mass, $\ev{m_{\beta\beta}}$, which quantifies the rate of this rare process is defined as
  \begin{equation}\label{mbb}
 \ev{m_{\beta\beta}}=|\sum_i U_{ei}^2 m_i|=|c_{12}^2c^2_{13}m_1 + s_{12}^2c_{13}^2e^{i\alpha_2}m_2 + s_{13}^2e^{i(\alpha_3-2\delta)}m_3|.
 \end{equation}
  We can calculate $\ev{m_{\beta\beta}}$ since all the quantities appearing in eq.(\ref{mbb}) can be calculated as a function of mixing angles and mass-squared differences. The plot of $\ev{m_{\beta\beta}}$
  as a function of $\theta_{23}$ is given in Fig.\ref{mbbvst23} along with the $n\sigma$ uncertainty bands
  arising form the $n\sigma$ errors in the measurement of mass-squared differences and mixing angles. The strongest upper bound on $\ev{m_{\beta\beta}}$ comes from the KamLAND-Zen experiment\cite{KamLAND-Zen:2016pfg} which restricts $\ev{m_{\beta\beta}}<0.061$eV. The plot of $\ev{m_{\beta\beta}}$ given in Fig.\ref{mbbvst23} and the plot of $m_3$ given in Fig.\ref{m3vst23} are very similar because of the quasi-degenerate hierarchy in our model. The best-fit ($1\sigma,~2\sigma,~3\sigma$) value(ranges) of $\ev{m_{\beta\beta}}$ is given by  $0.04((0.03-0.075),~(0.027-0.24),~(0.026-\infty))$. Due to the symmetry about $\theta_{23}=\pi/4$, the measurement of $\sum m$ and $\ev{m_{\beta\beta}}$ cannot resolve the octant ambiguity of $\theta_{23}$.  

The predictions for best-fit value and $n\sigma$ ranges of $\delta_{CP}$, $\alpha_{2,3}$, $m_3$, $\sum m$, and $\ev{m_{\beta\beta}}$ using the values of mixing angles and mass-squared differences from the Table.\ref{data} are summarized in the Table.\ref{predictions}.

 Analogous to the type C two-zero texture in our model, the  $L_\mu-L_\tau$ gauge symmetric models which employ type-I seesaw mechanism satisfy a two-zero minor texture allowing similar predictions\cite{Asai:2017ryy,Biswas:2016yjr,Biswas:2016yan}. The two-zero minor textures where two zeroes are present in the inverse of the light neutrino mass matrix is studied in \cite{Lavoura:2004tu,Lashin:2007dm,Verma:2011kz,Liao:2013saa}. The type-C two-zero minor pattern arising in the minimal $L_\mu-L_\tau$ gauge symmetric type-I model is studied in great detail in \cite{Asai:2017ryy}. The relations between the parameters, in this case, can be obtained by merely replacing the mass ratios $m_{2,3}/m_1$ with their reciprocals in eq.(\ref{massratios}). Complementary to our model, the type-I model only accommodates normal ordering with a quasi-degenerate hierarchy. Because of quasi-degenerate hierarchy in both models, the predictions are similar and most of the differences come from the different experimentally measured values of mixing angles and mass-squared differences for normal and inverted orderings. Unlike our model, the best-fit value of $\delta_{CP}$ cannot be accommodated even at $2\sigma$ range. The mass of the lightest neutrino is also larger in the inverse seesaw model allowing a higher sensitivity to the bounds from cosmology and neutrinoless beta decay. Therefore finalizing the neutrino hierarchy, a more precise measurement of $\delta_{CP}$, or stronger bounds from cosmology and neutrinoless beta decay can distinguish between these two models.

\begin{table}
\hskip-1.0cm\begin{tabular}{cc c c c}
\hline\hline
%------------------------------------------------------------------------------------------------------------------------------------------------------------
%------------------------------------------------------------------------------------------------------------------------------------------------------------
Parameter             &       best-fit      &     $1\sigma$                       &            $2\sigma$                &   $3\sigma$   \\
\hline\\
%------------------------------------------------------------------------------------------------------------------------------------------------------------
$\delta_{CP}/\pi$     &                     &  $(0.62-0.76)\oplus(0.3-0.4) $      &$(0.55-0.9)\oplus(0.2-0.5)$&                          \\
                      &      1.31,~0.69     &  $~~~~~~~~~~~~~~\oplus~~~~~~~~~~~$  &$~~~~~~~~~~~~~\oplus~~~~~~~~~~~$    &(0-2)           \\
                      &                     &  $(1.24-1.38)\oplus(1.6-1.7)$       &$~~~(1-1.45)\oplus(1.5-1.8)$            &                \\
\\
$\alpha_2/\pi$        &    $\pm 0.65$       &  $\pm(0.4-0.75)$                    &$\pm (0.16-0.9)$                     &(-1,1)          \\
\\
$\alpha_3/\pi$        &    $\pm 0.49$       &  $\pm(0.35-0.7)$                    &$\pm (0.1-0.9)$                     &(-1,1)           \\
\\
$m_3$                 &    $0.04$           &  $(0.03-0.07)$                      &$(0.027-0.24)$                      &$(0.026-\infty)$ \\
\\
$\sum m$              &    $0.169$          &  $(0.153-0.248)$                    & $(0.141-0.71)$                     &$(0.133-\infty)$ \\
\\
$\ev{m_{\beta\beta}}$ &    $0.04$           &  $(0.03-0.075)$                     &$(0.027-0.24)$                      &$(0.026-\infty)$ \\
%--------------------------------------------------------------------------------------------------------------------------------------------------------------
%--------------------------------------------------------------------------------------------------------------------------------------------------------------             
\hline\hline\\
\end{tabular}
\caption{The predictions for the values of CP-violating Dirac phase $\delta_{CP}$, Majorana phases $\alpha_{2,3}$, mass of the lightest neutrino, $m_3$, sum of neutrino masses $\sum m$, and the effective Majorana mass of neutrinoless double beta decay $\ev{m_{\beta\beta}}$ in the best-fit, $1,~2,~3\sigma$ ranges using the values of mixing angles and mass-squared differences in Table.\ref{data}.}
\label{predictions}
\end{table}

 %%%%%%%%%%%%%%%%%%%%%%%%%%%%%%%%%%%%%%%%%%%%%%%%%%%%%%%%%%%%%%%%%%%%%%%%%%%%%%%%%

%%%%%%%%%%%%%%%%%%%%%%%%%%%%%%%%%%%%%%%%%%%%%%%%%%%%%%%%%%%%%%%%%%%%%%%%%%%%%%%%%%
\section{Conclusion}

By combining three popular ideas in model building, viz., the L-R symmetry, inverse seesaw mechanism, and $L_\mu-L_{\tau}$ gauge symmetry, we propose the $G_{L-R}\times U(1)_{\mu-\tau}$ gauge symmetric model. The spontaneous symmetry breaking of the $U(1)_{L_\mu-L_\tau}$ allows us to resolve the discrepancy in the measurement of muon's anomalous magnetic moment and the longstanding puzzles of neutrino masses and oscillations. The best-fit point in the parameter space which accommodates the experimentally measured values of neutrino oscillation parameters and muon anomalous magnetic moment ($\Delta a_\mu$) is given in Table.\ref{fit}. The choice of inverse seesaw mechanism provides an added advantage of probing new physics from right-handed neutrinos in the multi-TeV scale.

 The presence of two zeros in the neutrino mass matrix from the $U(1)_{L_\mu-L_\tau}$ symmetry makes this model very predictive by providing four relations among the precisely measured values of mixing angles and mass-squared differences to make predictions for CP-violating Dirac phase, Majorana phases, and the absolute value of neutrino masses. These predictions along with their uncertainties coming from the error in the measurement of mixing angles are given in Table.\ref{predictions}. We see that in this model, only inverted ordering with a quasi-degenerate ordering is allowed. The predictions of complementary values of $\delta_{CP}$ in both octants of $\theta_{23}$ allow us to have octant sensitivity in our model. The current experimentally measured best-fit value and $1\sigma $ range of $\delta_{CP}$ mostly favor $\theta_{23}>\pi/4$. The prediction for $\delta_{CP}$ from the best-fit values of mixing angles and mass-squared differences surprisingly agrees with the current experimentally measured best-fit value. The large values of light neutrino masses also make this model very sensitive to cosmological bound on the sum of neutrino masses and the effective Majorana mass bounds from neutrinoless double beta decay. Most of the parameter space in the  $1\sigma$ range is allowed by these bounds as shown in Figs. \ref{summvst23} and \ref{mbbvst23}.
 \section{Acknowledgement}

I thank Prof. Rabindra Mohapatra for his invaluable help and discussions during the course of this research. 
\newpage
\appendix
\section{Gauge boson masses and mixing}\label{ES}

\begin{equation}\label{higgs-kinetic}
\mathcal{L}_{higgs-kinetic}= (D_\mu\chi_R)^\dagger(D_\mu\chi_R) + \text{Tr}[(D_\mu\phi)^\dagger(D_\mu\phi)] + (D_\mu\delta)^*(D_\mu\delta),
\end{equation}
where the covariant derivatives are 
\begin{eqnarray}
\notag
D_\mu\phi&=&\partial_\mu\phi - i\frac{g_L}{2}\vec{W}_{L\mu}.\vec{\sigma}\phi+i\frac{g_R}{2}\phi\vec{W}_{R\mu}.\vec{\sigma},\\ 
D_\mu\chi_R&=&\partial_\mu\chi_R - i\frac{g_R}{2}\vec{W}_{R\mu}.\vec{\sigma}\chi_R+i\frac{g_{B-L}}{2}B_\mu\chi_R,\\\notag
D_\mu\delta&=&\partial_\mu\delta - i \frac{g_{\mu\tau}}{2} B'_\mu \delta.\notag
\end{eqnarray}

The matrices $\vec{\sigma}$ are the three pauli matrices and $B$ and $B'$ are the gauge boson fields associated with $U(1)_{B-L}$ and $U(1)_{L_\mu-L_\tau}$ respectively. The gauge boson masses and mixing are determined by the SSB of $\mathcal{L}_{higgs-kinetic}$ given in the eq.(\ref{higgs-kinetic}).
Since $\phi$ transforms non-trivially under $SU(2)_L$ and $SU(2)_R$, the gauge eigen states $W_L$ and $W_R$ will mix to give mass eigen states $W$ and $W'$. Unlike the minimal L-R symmetric model, there is no mixing among the charged $W$ fields because of the vanishing VEV of $\phi_2^0$. This is not a general feature of the model but a simplification we enforced. The masses of charged gauge bosons after SSB are given as
\begin{eqnarray}
M_{W}^2&=&\frac{1}{2}\kappa^2g_L^2,\\ \notag
M_{W'}^2&=&\frac{1}{2}(\kappa^2 + v^2)g_R^2.
\end{eqnarray}
In the neutral gauge sector, we have four neutral gauge fields $W_{3R},~W_{3L},~B,$ and $B'$. Since $\delta$ is an L-R singlet, $B'$ does not mix with other gauge boson fields. After mixing, we get a massless photon field, $\gamma$, and three massive fields: $Z$, $Z'$, and $Z_{\mu\tau}$.

The masses of the massive neutral gauge bosons are calculated as
\begin{eqnarray}\label{gaugebosonmasses}\notag
&M_Z^2&=\frac{1}{2}\frac{\kappa^2(g_L^2g_R^2 + g_B^2(g_L^2 +g_R^2))}{g_B^2 + g_R^2}=\frac{M_{WL}^2}{\cos^2\theta_W},\\
&M_{Z'}^2&=\frac{1}{2}v^2(g_B^2 + g_R^2),\\\notag
&M_{Z\mu\tau}^2&=\frac{1}{2}g_\mu^2\delta^2.
% * <abhishdev92@gmail.com> 2017-03-20T06:30:32.869Z:
%
% ^.
\end{eqnarray}
The mass eigen states after mixing are given as
\begin{eqnarray}\notag
&\left(
\begin{array}{c}
 Z' \\
 Z \\
 \gamma  \\
\end{array}
\right)&= \left(
\begin{array}{ccc}
 0 & \cos\phi & -\sin\phi\\
 \cos\theta_W &-\sin \phi  \sin  \theta _W &-\cos\phi\sin\theta_W \\
 -\sin\theta_W & - \sin \phi   \cos  \theta _W & -\cos\phi\cos\theta_W \\
\end{array}
\right)\left(
\begin{array}{c}
 W_{3 L} \\
 W_{3 R} \\
 B\\
\end{array}
\right)~\text{and}\\
&Z_{\mu\tau}&=B',
\end{eqnarray}
where the Weinberg angle $\theta_W$ and the neutral mixing angle $\phi$ are given as
\begin{eqnarray}\notag
&\cos^2\theta_W&=\frac{g_L^2( g_B^2 + g_R^2)}{g_L^2g_R^2 +  g_B^2(g_L^2 + g_R^2)},\\
&\cos^2\phi &=\frac{g_R^2}{ g_B^2 + g_R^2}.
\end{eqnarray}

\section{Higgs potential}\label{HPappendix}

The most general gauge-invariant, left-right symmetric Higgs potential is given as
\begin{eqnarray}\notag\label{HP}
\mathcal{V}(\phi,\chi_R,\delta)=&-&\mu_1^2\delta\delta^* - \mu_2^2\chi^\dagger\chi-\mu_3^2\text{Tr}[\phi^\dagger\phi]-\mu_4^2[\text{Tr}(\tilde{\phi}^\dagger\phi) + \text{Tr}(\tilde{\phi}\phi^\dagger)]\\\notag
&+&\lambda_1 (\delta\delta^*)^2 + \lambda_2 (\chi^\dagger\chi)^2+\lambda_3 (\text{Tr}(\phi^\dagger\phi))^2+\lambda_4[(\text{Tr}(\tilde{\phi}\phi^\dagger))^2 + (\text{Tr}(\tilde{\phi}^\dagger\phi))^2]\\\notag
&+&\lambda_5\text{Tr}(\phi^\dagger\phi)[\text{Tr}(\tilde{\phi}\phi^\dagger)+\text{Tr}(\tilde{\phi}^\dagger\phi)]
+\lambda_6\text{Tr}(\tilde{\phi}\phi^\dagger)\text{Tr}(\tilde{\phi}^\dagger\phi)\\\notag
&+&\rho (\chi^\dagger\chi)(\delta \delta^*)\\\notag
&+&\alpha_1\text{Tr}(\phi^\dagger\phi)\chi^\dagger\chi + \alpha_2 \text{Tr}(\phi^\dagger \phi)(\delta\delta^*)\\\notag
&+& [e^{i\delta_3}\alpha_3 \text{Tr}(\tilde{\phi}^\dagger\phi)\chi^\dagger\chi + e^{i\delta_4}\alpha_4\text{Tr}(\tilde{\phi}^\dagger\phi)(\delta\delta^*)+ \text{h.c}]\\\notag
&+&\beta_1(\chi^\dagger\phi^\dagger\phi\chi) + \beta_2(\chi^\dagger\tilde{\phi}^\dagger\tilde{\phi}\chi)
+\beta_3(\tilde{\chi}^\dagger\phi^\dagger\phi\tilde{\chi})+\beta_4(\tilde{\chi}^\dagger\tilde{\phi}^\dagger\tilde{\phi}\tilde{\chi})\\
&+&[\beta_5 e^{i\delta_5}\chi^\dagger\tilde{\phi}^\dagger\phi\chi + \beta_6 e^{i\delta_6}\tilde{\chi}^\dagger\tilde{\phi}^\dagger\phi\tilde{\chi} + \text{h.c}],
\end{eqnarray}
where we have defined the left-right symmetry as the exchange $\phi\leftrightarrow\phi^\dagger,~\delta \leftrightarrow \delta^*,~\text{and}~\chi_R \leftrightarrow \chi_R^\dagger$. We have also used the compact notation $\chi$ for $\chi_R$ in eq.(\ref{HP}). All the couplings appearing in eq.(\ref{HP}) are real.

\newpage

\newpage

\bibliographystyle{unsrt}
\bibliography{mybib.bib}

\end{document}